\documentclass[12pt,preprint,apj]{emulateapj}
\usepackage{enumerate}
\usepackage{amsmath}

\usepackage{multirow}
\usepackage{bigstrut}

\usepackage{tablefootnote}

\shorttitle{High-speed multiple encounters of a late-type galaxy with early-type galaxies}
\shortauthors{Hwang et al.}

\begin{document}

\title{Evolution of late-type galaxies in a cluster environment: 
Effects of high-speed multiple encounters with early-type galaxies}
\author{Jeong-Sun Hwang$^{1}$, Changbom Park$^{2}$, Arunima Banerjee$^{3}$, and Ho Seong Hwang$^{4}$}
\affil{$^1$ 
Department of Physics and Astronomy, Sejong University, 209 Neungdongro, Gwangjin-gu, Seoul 05006, Korea; hwang2k@gmail.com}
\affil{$^2$ 
School of Physics, Korea Institute for Advanced Study, 85 Hoegiro, Dongdaemun-gu, Seoul 02455, Korea}
\affil{$^3$ 
Indian Institute of Science Tirupati, Tirupati {\textendash} 517507, India}
\affil{$^4$ 
Quantum Universe Center, Korea Institute for Advanced Study, 85 Hoegiro, Dongdaemun-gu, Seoul 02455, Korea; hhwang@kias.re.kr}

\begin{abstract} 
Late-type galaxies falling into a cluster would evolve 
being influenced 
by the interactions with both the cluster and the nearby cluster member galaxies. 
Most numerical studies, however, tend to focus on the effects of the former 
with little work done on those of the latter. 
We thus perform a numerical study on the evolution of a late-type galaxy 
interacting with neighboring early-type galaxies at high speed, 
using hydrodynamic simulations. 
Based on the information obtained from 
the Coma cluster, 
we set up the simulations for the case where a 
Milky Way-like late-type galaxy  
experiences six consecutive collisions with 
twice as massive early-type galaxies 
having hot gas in their halos  
at the closest approach distances of  
15{\textendash}65~$h^{-1}$~kpc
at the relative velocities of 
1500{\textendash}1600~km~s$^{-1}$.
Our simulations show 
that the evolution of the late-type galaxy can be significantly 
affected by the accumulated effects of the 
high-speed multiple collisions with
the early-type galaxies, 
such as on cold gas content and 
star formation activity of the late-type galaxy, 
particularly through the hydrodynamic interactions 
between cold disk and hot gas halos.  
We find that the late-type galaxy 
can lose most of its cold gas after the six collisions 
and have more star formation activity during the collisions. 
By comparing our simulation results with 
those of galaxy{\textendash}cluster interactions,  
we claim that the role of the galaxy{\textendash}galaxy interactions 
on the evolution of late-type galaxies in clusters 
could be comparable with that of the galaxy{\textendash}cluster interactions, 
depending on the dynamical history. 
\end{abstract}



\section{INTRODUCTION}

The evolution of galaxies is driven by both nature and nurture, 
and understanding the
relative importance between the two has been one of important issues in astrophysics 
(e.g. \citealp{Peng+2010}). 
It has been well known that  
the relative abundance of galaxies with different morphological types 
is closely related 
to the densities of their local environment 
(e.g. \citealp{Dressler1980}; \citealp{Postman+2005}; \citealp{Park+2007}; 
\citealp{Ann2017}; \citealp{LHuillier+2017}); 
the fraction of spiral galaxies is observed to be highest 
in the field ($\sim$80\%), followed by the periphery of clusters ($\sim$60\%) 
to almost nil ($\sim$0\%) in the centers of rich 
clusters (\citealp{Dressler1980}; \citealp{Whitmore+1993}). 
The other physical properties of cluster galaxies are also  
found to be 
starkly different from those of their 
field counterparts (e.g. \citealp{Mastropietro+2005}; \citealp{Park_Hwang2009}; 
\citealp{von der Linden+2010}; \citealp{Hwang+2012}; 
\citealp{Sheen+2017}; \citealp{Song+2017}). 
This possibly indicates that environment does play a significant role 
in regulating the structure and morphology 
and therefore the evolution of galaxies.

The motivation behind studying late-type galaxies (LTGs) 
in cluster environment may not be obvious. 
This is because LTGs are relatively fewer in number and are, 
therefore, not quite representative of the general galaxy population. 
This apart, they are  known to have evolved via the route of 
secular evolution, which effectively overrules environmental effects; 
hence late types are possibly not the ideal testbeds  
to study the effect of environment on galaxy evolution. 
However, LTGs are characterized by stellar 
and gaseous disks, which are cold, diffuse, and hosted in shallow 
gravitational potentials and are, hence fragile; therefore, they may serve as 
useful diagnostic tracers of environmental effects on galaxy properties 
(e.g. \citealp{Blanton_Moustakas2009}; \citealp{Hwang+2010}).
Besides, having evolved via secular evolution, 
they may reliably estimate the 
ages of the different galaxy components, 
unlike their early-type counterparts. 
Finally, there is an observed preponderance of the infall of LTGs 
into clusters in the current cosmological epoch, which makes it imperative 
to study the evolution of LTGs in a cluster environment.

As indicated earlier, the characteristics of LTGs in cluster environment 
are significantly different from those of 
their field counterparts (e.g. \citealp{Boselli_Gavazzi2006}). 
A deficiency in cold, neutral hydrogen (HI) is perhaps the most definite 
fingerprint of a rich cluster 
environment (\citealp{Haynes+1984}; \citealp{Giovanelli_Haynes1985}). 
In particular, LTGs  in cluster environment are found to have 
less extended or truncated HI disks, in addition to having signatures 
of sloshing and lopsidedness, which are characteristic of 
gas-poor galaxies in general. 
They are also marked by redder colors, lower rate of star formation, 
and truncated star-forming disks (\citealp{Koopmann+2006}). 
Interestingly, all these observational features appear within the 
X-ray-emitting gas or the hot halo of the cluster, thus underscoring 
the role of the hot halo in regulating the evolution 
of the LTGs  
hosted in it (\citealp{Gavazzi+2006}). 
The characteristics of LTGs in cluster environment 
as discussed above are regulated by various physical mechanisms, 
which are either gravitational or hydrodynamic in nature. 
While the gravitational interactions mainly culminate in the development 
of tidal features, the hydrodynamic interactions between the galactic 
interstellar medium (ISM) and the hot ambient medium 
may lead to ram-pressure stripping 
and quenching of star formation, 
among others (See, for example, \citealp{Binney_Tremaine1987}).

The possible role played by the ambient hot gas in regulating 
the structure and evolution of LTGs has been lately 
analyzed in several numerical and observational studies 
(e.g., \citealp{Abadi+1999}; \citealp{Schulz_Struck2001}; 
\citealp{Vollmer+2001}; \citealp{Jachym+2007}; \citealp{Smith+2013}). 
But these earlier studies generally focused on the effect of 
hot gas associated with the intracluster medium (ICM) on 
the ISM of the LTGs. 
However, invoking the physics of just the ambient ICM  could 
not always satisfactorily explain certain observed features in the ISM. 
For instance, some member LTGs in clusters show 
characteristic signatures of ram-pressure stripping; however, 
the direction of ram pressure appears uncorrelated with 
the direction to the cluster center, 
which is one of the primary requirements for the ram-pressure stripping 
to be effective (e.g., see Figure 2 of \citealp{Ebeling+2014}). 
This already indicates that the effect of just the ambient ICM 
is not enough to understand the complex features observed 
in the ISM of member LTGs in the cluster environment.

Interestingly, X-ray observations of galaxy clusters have revealed that 
some early-type galaxies (ETGs) in clusters 
possess substantial 
hot gaseous halos, and their role in regulating the ISM of 
the neighborhood LTGs was reasonably promiscuous. 
The case of NGC~4438{\textendash-}an LTG in the Virgo Cluster 
in the neighborhood of M86, a giant elliptical in the neighborhood 
of hot halo gas{\textendash-}could be a good example.
In fact, \citet{Ehlert+2013} argued that the hot gas present 
in the halo of M86 strongly regulates the ISM of NGC~4438, 
while both are undergoing ram-pressure stripping 
by the ICM.  
Besides, \citet{Vollmer2009} showed that 
NGC~4388 and NGC~4438, the two LTGs near M86, 
require several times higher peak ram pressure than 
expected from a smooth and static ICM, 
using the dynamical models and X-ray observations. 
This again prompts the need to invoke physics beyond that of 
the ambient ICM in understanding the ISM of member LTGs 
in a cluster environment.

Moreover, \citet{Park_Hwang2009} 
have shown that  
galaxy{\textendash}galaxy encounters 
can strongly affect the properties of cluster LTGs 
through gravitational/hydrodynamic interactions 
of the LTGs with nearby ETGs 
by analyzing the Sloan Digital Sky Survey 
(SDSS; \citealp{York+2000}) galaxies 
associated with the Abell clusters.
They have found that the hydrodynamic interactions between the members 
play a dominant role in star formation quenching, 
while the hot cluster gas  
plays a relatively minor role. 
They also stressed 
the effects of both galaxy{\textendash}galaxy hydrodynamic 
interactions and galaxy{\textendash}cluster/galaxy{\textendash}galaxy gravitational interactions 
on the morphology transformation of the LTGs in clusters.

Motivated by such observational evidence highlighting 
the effects of galaxy{\textendash}galaxy interactions 
on cluster galaxy properties,  
this work aims to study the evolution of LTGs via 
interaction with early-type cluster member galaxies 
using numerical simulations. 
We focus on the cases of 
an LTG, flying either edge-on or face-on, 
experiencing high-speed multiple collisions with 
neighboring ETGs that contain hot gas in their surrounding halos, 
in order to investigate the 
effects of hydrodynamic interactions of cold disk gas and hot halo gas.
We examine the variation of the LTG properties, 
such as cold gas content and star formation activity, 
while undergoing consecutive collisions.
To assess the influence of both 
the hot halo gas of the colliding ETGs  
and the hot cluster gas 
on the evolution of LTGs,   
we compare the results of our simulations  
with those of galaxy{\textendash}cluster interactions 
in comparable simulation settings.

This paper is organized as follows. 
In Section 2, we present our galaxy models; 
in Section 3, our simulation code; and  
in Section 4, the initial setup of the encounters. 
In Section 5, we  discuss the results 
showing the evolution of an LTG in a cluster environment. 
Finally, we present the summary and discussion in Section 6.


\section{GALAXY MODELS}

For this numerical study, 
we construct an LTG model ``L" and 
an ETG model ``EH" using the 
ZENO\footnote{http://www.ifa.hawaii.edu/$\sim$barnes/software.html} 
software package (\citealt{Barnes2011}).  
We generate the models following the procedures described in 
\citet{Hwang+2013} and \citet{Hwang_Park2015}. 
Both models~L and EH for the current work 
adopt the same density models for all components (i.e., bulge, halo, and/or disk) 
as those used for the LTG and ETG models 
in \citet{Hwang_Park2015}, respectively, 
with minor changes in some model parameters.   
The key parameter values of the models 
are summarized in Table~1. 
In the following subsections, we give an overview of our models.
(As shown in Appendix~A, 
we have checked the stability of our models 
by evolving each model for several Gyr in isolation.)

\begin{deluxetable*}{llcc}
\centering
\tablecolumns{4}
\tablewidth{0pc}
\footnotesize
\tablecaption{Initial galaxy models \label{tab02}}
\tablehead{
\colhead{ } &
\colhead{ } &
\colhead{Model L} &
\colhead{Model EH}
}
\startdata
$M_{\rm tot}$ ($10^{10}~h^{-1}$ $M_{\odot}$) & 
  Total mass of the system 
       & 88.9 &   177.8\\
$R_{\rm vir}$\tablenotemark{a} ($h^{-1}$~kpc) & Virial radius 
       & 150 &   175\\       
 $f_{\rm dg}$\tablenotemark{b} & Disk gas fraction
    & 0.13  & $\cdots$  \\
 $f_{\rm hg}$\tablenotemark{c} & Halo gas fraction
    & $\cdots$   & 0.01 \\
\hline
\textbf{Gas disk:}\\
Disk model &   &  Exponential  &   $\cdots$ \\
$a_{\rm dg}$ ($h^{-1}$~kpc) & Gas disk scale length      & 6.125
                                                    &   $\cdots$ \\
$z_{\rm dg}$ ($h^{-1}$~kpc) & Vertical disk scale height    & 0.245   &   $\cdots$ \\
$b_{\rm dg}$ ($h^{-1}$~kpc) & Outer disk cutoff radius & 73.5   &   $\cdots$ \\
$M_{\rm dg}$ ($10^{10}~h^{-1}$ $M_{\odot}$) & Total gas disk mass   & 0.56   &   $\cdots$ \\
$N_{\rm dg}$  & Number of particles   & 32 768 &       $\cdots$ \\
$m_{\rm dg}$ ($10^{5}~h^{-1}$ $M_{\odot}$) & Mass of individual particles
                                              & 1.71
                                              & $\cdots$
                                              \\
$\epsilon_{\rm dg}$ ($h^{-1}$~kpc) & Gravitational softening length  &
                             0.077  & $\cdots$ \\
\hline
\textbf{Star disk:}\\
Disk model &   & Exponential  & $\cdots$  \\
$a_{\rm ds}$ ($h^{-1}$~kpc) & Star disk scale length & 2.45  & $\cdots$ \\
$z_{\rm ds}$ ($h^{-1}$~kpc)            & Vertical disk scale height    & 0.245 &   $\cdots$\\
$b_{\rm ds}$ ($h^{-1}$~kpc) & Outer disk cutoff radius & 29.4 &    $\cdots$ \\
$M_{\rm ds}$ ($10^{10}~h^{-1}$ $M_{\odot}$)  & Total star disk mass   & 3.64  & $\cdots$ \\
$N_{\rm ds}$       & Number of particles  &  122 880   & $\cdots$ \\
$m_{\rm ds}$ ($10^{5}~h^{-1}$ $M_{\odot}$) & Mass of individual particles
                                          & 2.96
                                          & $\cdots$
                                               \\
$\epsilon_{\rm ds}$ ($h^{-1}$~kpc) & Gravitational softening length  &
                             0.098 &  $\cdots$ \\
\hline
\textbf{Bulge:} \\
Bulge model & &  Hernquist &  Hernquist\\
$a_{\rm b}$ ($h^{-1}$~kpc)  & Bulge scale length   & 0.49 &   1.96 \\
$b_{\rm b}$ ($h^{-1}$~kpc) & Truncation radius  & 98 &  196 \\
$M_{\rm b}$ ($10^{10}~h^{-1}$ $M_{\odot}$)  & Total bulge mass   & 0.7  &  9.8 \\
$N_{\rm b}$ & Number of particles
          & 24 576   &   344 064 \\
$m_{\rm b}$ ($10^{5}~h^{-1}$ $M_{\odot}$) & Mass of individual particles
                                    & 2.85
                                & 2.85
                                    \\
$\epsilon_{\rm b}$ ($h^{-1}$~kpc) & Gravitational softening length  &
                               0.098 &  0.098
                               
                                \\
\hline                                  
\textbf{Gas halo:} \\
Halo model &   &  $\cdots$   &   Isothermal\\
$a_{\rm hg}$ ($h^{-1}$~kpc)
              & Core radius &  $\cdots$   &   8.4   \\
$b_{\rm hg}$ ($h^{-1}$~kpc)
             & Tapering radius   &  $\cdots$ &  252  \\
$M_{\rm hg}$ ($10^{10}~h^{-1}$ $M_{\odot}$) & Total gas halo mass
		&  $\cdots$   &  1.68 \\
$N_{\rm hg}$ & Number of particles
        &  $\cdots$   &  98 304 \\
$m_{\rm hg}$  ($10^{5}~h^{-1}$ $M_{\odot}$) & Mass of individual particles
                                            &  $\cdots$
                                              & 1.71
                                            \\
$\epsilon_{\rm hg}$ ($h^{-1}$~kpc)  & Gravitational softening length  &
                              $\cdots$  &  0.077 \\
\hline
\textbf{DM halo:} \\
halo model & & NFW  &   NFW\\
$a_{\rm hd}$ ($h^{-1}$~kpc)       & DM halo scale length  &   14    &   17.5  \\
$b_{\rm hd}$ ($h^{-1}$~kpc)        & Tapering radius  & 42 &   52.5   \\
$M_{\rm hd}(a_{\rm hd})$ ($10^{10}~h^{-1}$ M$_{\odot}$) & Mass within radius $a_{\rm hd}$
                                                  & 9.21  &  18.24      \\
$M_{\rm hd}(\infty)=M_{\rm hd}$ ($10^{10}~h^{-1}$ M$_{\odot}$) & Total DM halo mass  &
        84  &  166.32 \\
$N_{\rm hd}$ & Number of particles
          & 655 360  &   1 310 720 \\
$m_{\rm hd}$ ($10^{5}~h^{-1}$ $M_{\odot}$) & Mass of individual particles
                                            & 12.82 &
                                              12.7  \\
$\epsilon_{\rm hd}$ ($h^{-1}$~kpc)  & Gravitational softening length  &
                             0.21 &  0.21  
\enddata
\tablenotetext{a}{
We use the virial radius $R_{\rm vir}$ as $R_{200}$, which is defined as the radius within which the average density is 200 times the critical density.}
\tablenotetext{b}{$f_{\rm dg} = M_{\rm dg}/(M_{\rm dg} + M_{\rm ds})$.}
\tablenotetext{c}{$f_{\rm hg} = M_{\rm hg}/(M_{\rm hg} + M_{\rm hd})$.}
\end{deluxetable*}


\subsection{Model~L}

The LTG model~L is a Milky Way{\textendash}like model. 
The total mass of the model system is set to  
$88.9 \times 10^{10}\,h^{-1}{M_{\odot}}$  
(Table~1; $h$ is set to 0.7 throughout this paper). 
The virial radius of the model is $\sim$150~$h^{-1}$~kpc. 
It consists of the four 
components{\textendash-}a stellar disk, a gaseous disk, 
a stellar bulge, and a dark matter (DM) halo.

Both star and gas disks follow  
an exponential surface density profile 
and a sech$^2$ vertical profile, with radial and vertical 
scale lengths of $a_{\rm dc}$ and $z_{\rm dc}$, respectively 
(where the subscript ``$\rm d$" stands for ``disk" 
and the subscript ``$\rm c$" stands for 
either ``$\rm s$" for the star disk component 
or ``$\rm g$" for the gas disk component):  
\begin{equation}
  \rho_{\rm dc}(R,z) =
    \frac{M_{\rm dc}}{4 \pi a_{\rm dc}^2 z_{\rm dc}} \,
      e^{- R / a_{\rm dc}} \,
        \mathrm{sech}^2 \left( \frac{\it z}{\it z_{\rm dc}} \right) \, ,
  \label{eq1}
\end{equation}
The radial scale lengths 
of the star and gas disks 
are set to $a_{\rm ds}$ = 2.45~$h^{-1}$~kpc and 
$a_{\rm dg}$ = 2.5~$a_{\rm ds}$, respectively, 
and the vertical scale length of both disks is chosen to 
$z_{\rm ds}$ = $z_{\rm dg}$ = 0.1~$a_{\rm ds}$ (cf. \citealt{Moster+2011}). 
The total masses of the star and gas disks are 
$M_{\rm ds} = 3.64 \times 10^{10}\,h^{-1}{M_{\odot}}$ and 
$M_{\rm dg} = 0.56 \times 10^{10}\,h^{-1}{M_{\odot}}$, respectively.
Thus, the gas fraction in the disk, 
defined as $f_{\rm dg} = M_{\rm dg}/(M_{\rm ds} + M_{\rm dg})$, 
is about 0.13.
The surface density of the gas disk is shown in Figure~1 (left panel). 
Both disks rotate in clockwise direction. 
The gas particles on the disk move with the local circular velocities, 
while the stellar particles have additional velocity dispersions 
to the circular velocities as described in \citet{Barnes_Hibbard2009}. 
The temperatures of the disk gas particles  
are set to the single value of $T = 10,000$~K at the initial time 
(cf. top left panel of Figure~2).

The bulge component, which consists of stars only, 
follows the Hernquist profile (\citealt{Hernquist1990}) 
with truncation at large radii:  
\begin{equation}
  \rho_{\rm b}(r) =
    \left\{
      \begin{array}{ll}
	\displaystyle
          \frac{a_{\rm b} M_{\rm b}}{2 \pi} \,
	      \frac{1}{r (a_{\rm b} + r)^{3}} \, &
	     {\rm for} \,\, r \le b_{\rm b} \, , \\ [0.4cm]
	\displaystyle
	  \rho_{\rm b}^{*} \, \left(\frac{b_{\rm b}}{r}\right)^2 \,
	      e^{-2 r / b_{\rm b}} \,  &
	     {\rm for} \,\, r > b_{\rm b} \, . \\
      \end{array}
    \right.
  \label{eq2}
\end{equation} 
The radial scale length and the truncation radius 
are set to $a_{\rm b} = 0.49$   
and $b_{\rm b}$ = 98~$h^{-1}$~kpc, respectively (\citealt{McMillan_Dehnen2007}). 
The total mass of the bulge is chosen to 
$M_{\rm b} = 0.7 \times 10^{10}\,h^{-1}{M_{\odot}}$. 
Then, the total mass of the two stellar 
components (stellar disk and stellar bulge) becomes 
$M_{\rm ds} + M_{\rm b}  = 4.34 \times 10^{10}\,h^{-1}{M_{\odot}}$, 
in good agreement with recent observations of 
the Milky Way (e.g., \citealt{McMillan2011}; 
\citealt{Licquia_Newman2015}).

The DM halo follows a \cite{Navarro+1996} model 
known as a Navarro{\textendash}Frenk{\textendash}White (NFW) profile, 
with an exponential taper at rage radii: 
\begin{equation}
\rho_{\rm hd}(r) =
  \left\{
  \begin{array}{ll}
   \displaystyle
   \frac{M_{\rm hd}( a_{\rm hd})}{4 \pi (\ln(2) - \frac{1}{2})} \,
   \frac{1}{r (r + a_{\rm hd})^2} &
	    {\rm for} \,\, r \le b_{\rm hd} \, , \\ [0.4cm]
   \displaystyle
   \rho_{\rm hd}^{*} \, \left(\frac{b_{\rm hd}}{r}\right)^2 \,
	    e^{-2 \beta (r / b_{\rm hd} -1)} &
	    {\rm for} \,\, r > b_{\rm hd} \, . \\
  \end{array}
  \right.
\label{eq3}
\end{equation}
The radial scale length and the tapering radius are chosen to  
$a_{\rm hd}$ = 14 and $b_{\rm hd}$ = 42~$h^{-1}$~kpc, respectively.
The total mass of the DM component is set to 
$M_{\rm hd}$ = 84~$\times 10^{10}\,h^{-1}{M_{\odot}}$.

\begin{figure} [!hbt]
\centering
\includegraphics[width=8.0cm]{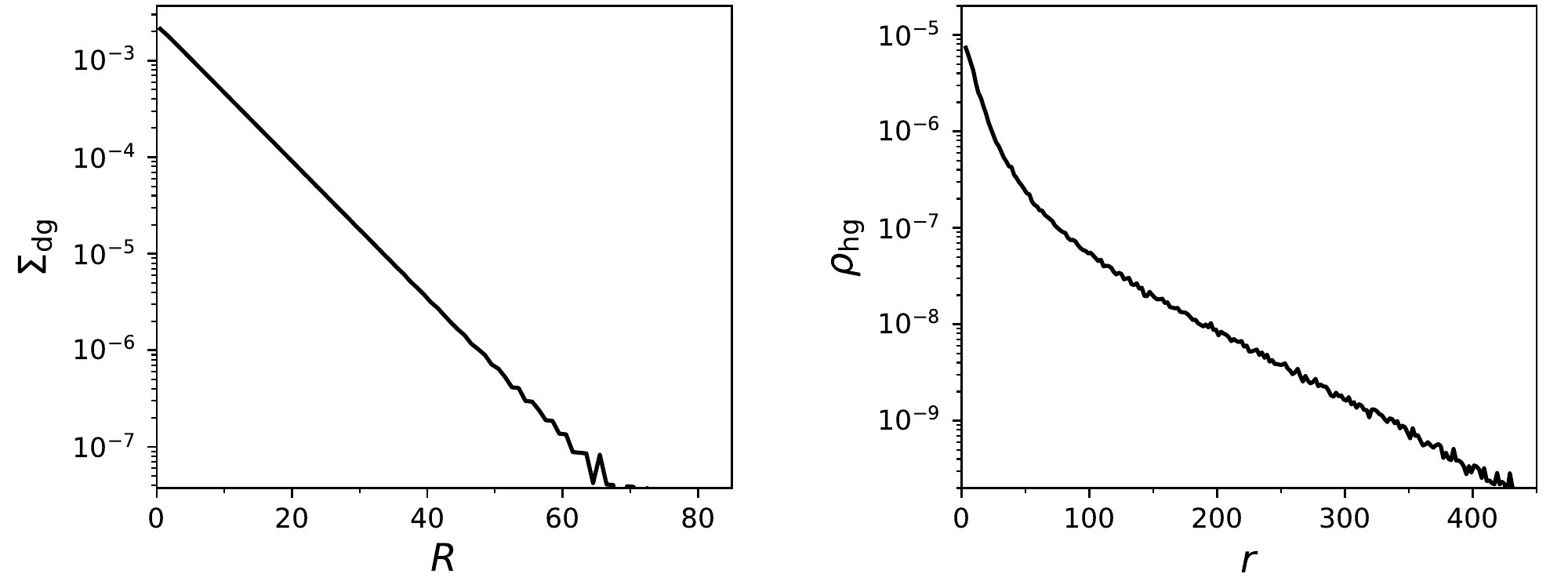}
\caption{
{{Left:}} surface density profile of 
the gas disk of model~L. 
The surface density $\Sigma_{\rm dg}$ is in the unit of 
$10^{10}\,h\,{M_{\odot}}\,{\rm{kpc}}^{-2}$, 
and the cylindrical radius $R$ is in $h^{-1}$~kpc. 
{{Right:}} spherically averaged density profile of 
the gas halo of model~EH. 
The gas density $\rho_{\rm{hg}}$ is in the unit of 
$10^{10}\,h^{2}{M_{\odot}}\,{\rm{kpc}}^{-3}$, 
and the spherical radius $r$ is in $h^{-1}$~kpc.  
}
\end{figure}


\subsection{Model~EH}

The ETG model~EH is intended to be twice as massive as 
the LTG model~L.  
The total mass and virial radius of model~EH are 
$177.8 \times 10^{10}\,h^{-1}{M_{\odot}}$ 
and $\sim$175~$h^{-1}$kpc, respectively (Table~1).
Model~EH is composed of a stellar bulge, DM halo, 
and gaseous halo.

As in model~L, the stellar bulge component 
has the Hernquist profile (Equation~(2)) 
with a radial scale length of $a_{\rm b}$ = 1.96~$h^{-1}$~kpc. 
The total mass is $M_{\rm b} = 9.8 \times 10^{10}\,h^{-1}{M_{\odot}}$, 
which is two times the total disk-plus-bulge mass of model~L.

The DM halo component also follows the NFW profile (Equation~(3)) 
as in model~L, 
with a length scale and the tapering radius of 
$a_{\rm hd}$ = 17.5 and 
$b_{\rm hd}$ = 52.5~$h^{-1}$~kpc, respectively. 
The total mass of the DM halo is 
$166.32 \times 10^{10}\,h^{-1}{M_{\odot}}$.

Unlike in model~L, 
a gas halo component is included in model~EH.  
The gas halo follows an isothermal profile with truncation:  
\begin{equation}
	\rho_{\rm hg}(r) =
	\displaystyle
	\frac{f_{\rm norm}  M_{\rm hg}}{2 {\pi} \sqrt{\pi} \, b_{\rm hg}} \,
	\frac{1}{r^2 + a_{\rm hg}^2} \,  e^{- (r / b_{\rm hg})^2}
	\, .  \\
	\label{eq4}
\end{equation}
The core radius is set to $a_{\rm hg}$ = 8.4~$h^{-1}$~kpc, and 
the tapering radius is chosen to be $b_{\rm hg}$ = 252~$h^{-1}$~kpc. 
The total mass of the gas halo is 
$M_{\rm hg} = 1.68 \times 10^{10}\,h^{-1}{M_{\odot}}$. 
The halo gas fraction, 
$f_{\rm hg} = M_{\rm hg}/(M_{\rm hd} + M_{\rm hg})$, 
is 0.01.   
The radial density profile of the gas halo is 
presented in Figure~1 (right panel). 
The initial temperatures of the gas halo particles are 
determined by the hydrostatic equilibrium 
(cf. top row second left panel in Figure~2).

\begin{figure*} [!hbt]
\centering
\includegraphics[width=13.0cm]{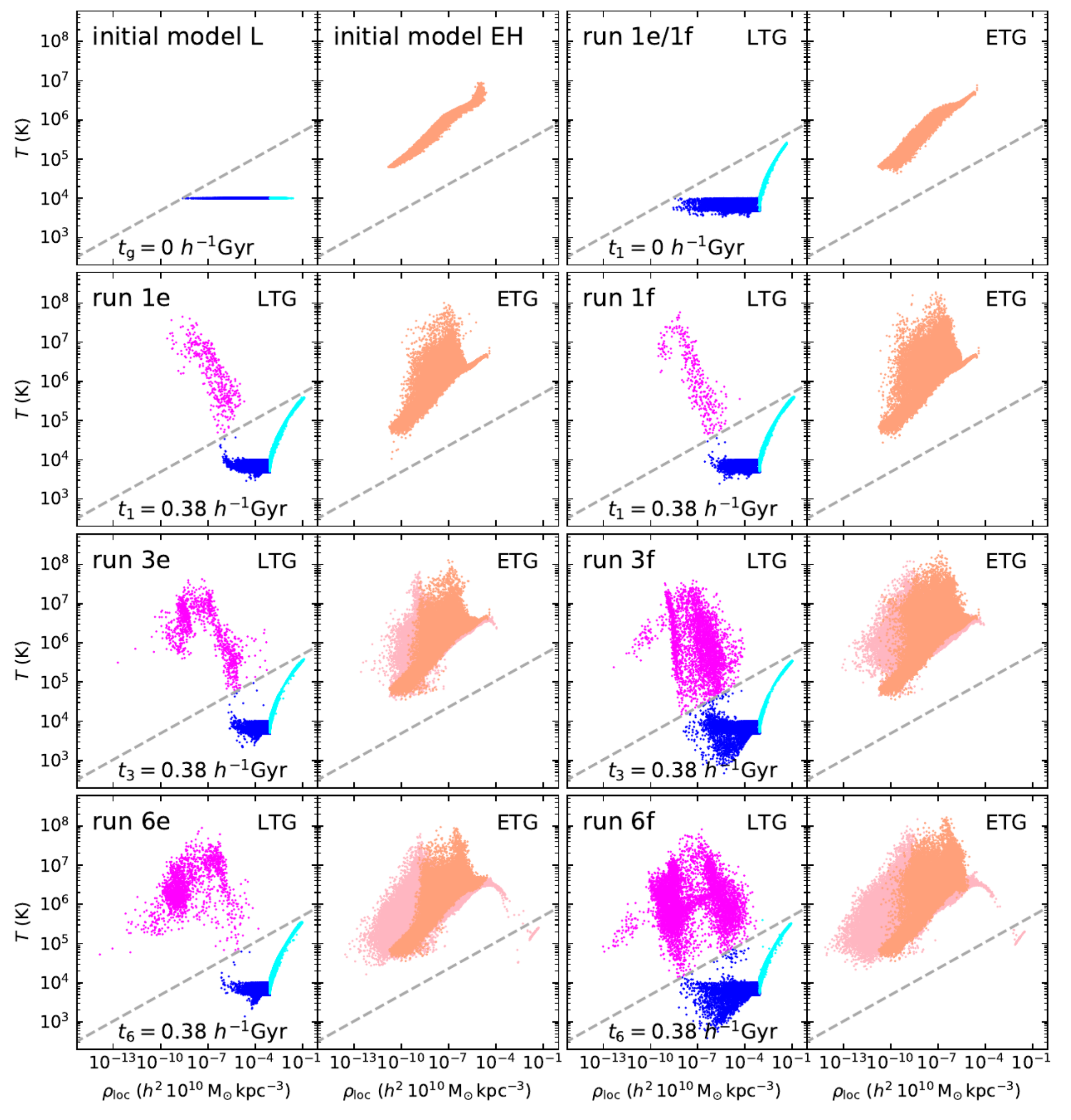}
\caption{
Temperature ($T$) against local density ($\rho_{\rm loc}$) 
for the gas particles of the LTG and ETG models. 
{{Two top left panels:}} 
properties of the disk gas of model~L (left panel) and 
the halo gas of model~EH (right panel) in isolation 
at $t_{\rm g} = 0$,   
i.e., at the time of the realization of each model. 
The disk gas and halo gas are well separated from 
each other in the $\rho_{\rm loc}${\textendash}$T$ 
plane by the dashed line. 
Among the disk gas, the star-forming gas is shown in cyan, 
and the non-star-forming ``cold" gas (lying below the dashed line) is 
shown in blue. 
The ``hot" halo gas (lying above the dashed line) of model~EH 
is plotted in light brown. 
{{Two top right panels:}} 
gas properties at $t_{1} = 0$, i.e., at the start of runs~1e and 1f.  
Models~L and EH$_1$ included at this time step 
for the first-encounter simulations 
are those evolved in isolation for 0.7~$h^{-1}$~Gyr. 
The disk gas of model~L and the halo gas of model~EH$_1$ 
are still well separated by the dashed line.  
{{Remaining rows:}} 
gas properties shortly (0.03~$h^{-1}$~Gyr) after the 
first, third, and sixth collisions (second, 
third, and fourth rows, respectively) 
in the edge-on (left two columns) and face-on (right two columns) cases. 
(Here $t_{i}$ with ${i}$ = 1, 2, ..., 6 represents 
the time elapsed since the start of the 
$i$th encounter simulations. 
The collisions between the LTG and the ETG models   
in the $i$th-encounter simulations occur 
at $t_{i} = 0.35$~$h^{-1}$~Gyr, refer to Table~2.) 
Some of the particles originally set as the disk gas of model~L 
have crossed the dashed line and become hot gas (magenta).    
In the third row, 
the halo gas particles of models~EH$_{1}$ and EH$_{2}$ are plotted 
first (pink), and then those of model~EH$_{3}$ are overplotted (light brown). 
Similarly, in the fourth row,  
the halo gas particles of model~EH$_{6}$ (light brown) are plotted on top of 
those of models~EH$_{1}$ through EH$_{5}$ (pink).  
In the bottom two rows, 
many more hot gas particles of model~L (magenta) are seen 
in the face-on cases than in the edge-on cases. 
}
\end{figure*}


\section{Simulation code}

For the simulations of the galaxy{\textendash}galaxy interactions, 
we use an early version of 
the $N$-body/smoothed particle hydrodynamics (SPH) code GADGET-3 
(originally described in \citealt{Springel2005}),  
the same version of the code as we used 
in \citet{Hwang_Park2015}. 
Here we briefly describe the simulation code and 
refer interested readers 
to \citet{Springel_Hernquist2003} and \citet{Hwang_Park2015} 
for a more detailed description.

The code uses a tree algorithm (\citealt{Barnes_Hut1986}) 
for calculating the gravitational force. 
For computing the hydrodynamic force,  
the SPH method 
in the entropy conservative formulation 
is adopted with a spline kernel 
(\citealt{Gingold_Monaghan1977}; \citealt{Springel_Hernquist2002}).
The radiative cooling and heating are taken into account  
for the primordial mixture of hydrogen and helium 
(\citealt{Katz+1996}).
Star formation and supernova feedback in the ISM  
are also implemented using the effective multiphase model of 
\citet{Springel_Hernquist2003}.

For the parameters related to star formation and feedback, 
we adopt the standard values of the multiphase model 
in all of our simulations.  
The star formation time-scale is set to  
$t_{0}^{\star} = 1.5~h^{-1}$~Gyr. 
The mass fraction of massive stars is $\beta = 0.1$.
The ``supernova temperature" and the temperature of cold clouds 
are $T_{\rm SN} = 10^{8}$ and 
$T_{\rm c} = 1000$~K, respectively. 
The parameter value for supernova evaporation is $A_{0} = 1000$.

We set the gravitational softening lengths for the particles  
in such a way that 
the maximum acceleration experienced by a single particle 
is equal in each component. 
Specifically, the softening lengths for the gas (both halo and disk gas), 
DM, disk star, and bulge particles 
are set to 
0.077, 0.21, 0.098, and 0.098~$h^{-1}$~kpc, respectively (Table~1).


\section{Initial setup of the encounters}

The aim of our numerical study is to examine 
how an LTG (target galaxy) falling into a cluster 
evolves, particularly through multiple encounters with 
cluster ETGs possessing hot halo gas. 
In order to construct the initial conditions (ICs) of our simulations   
for plausible cases of the interactions, 
we use the information about 
the spatial distribution of galaxies drawn 
from the galaxy catalog of the Coma  
cluster (H.~S. Hwang et al. 2018, in preparation).

In the following subsections, 
we first explain the Coma cluster catalog and 
the way we estimate the three-dimensional (3D) volume 
density of the member galaxies. 
Then, we describe the procedure to build the ICs 
for the consecutive collisions between our LTG and ETG models.


\subsection{Galaxies in the Coma cluster}

The Coma is a well-known nearby cluster at $z$=0.023.
The estimated mass, radius, and the velocity dispersion of 
the cluster are 
$M_{200} = 1.29^{+0.15}_{-0.15} \times 10^{15}\,{M_{\odot}}$, 
$R_{200} = 2.23^{+0.08}_{-0.09}$~Mpc, 
and $\sigma_{\rm cl} = 947 \pm 31$~km~s$^{-1}$, 
respectively (\citealp{Sohn+2017}).

We have conducted a new redshift survey of the Coma cluster  
to uniformly and densely cover the cluster region with MMT/Hectospec. 
We did not use any color selection criteria for spectroscopic targets. 
By combining with the existing SDSS data 
in this region (mainly $r$ $<$ 17.77), 
we could increase the magnitude limit for the redshift data 
up to $r$ = 20. 
Among 4761 galaxies with measured redshifts within the
radius of 1.9~$h^{-1}$Mpc, 
we use 1088 member galaxies identified with the 
caustic technique 
(\citealp{Diaferio_Geller1997}; \citealp{Diaferio1999}).
The catalog includes right ascension,  
declination, and projected clustercentric radius, 
stellar mass, morphological type, 
Petrosian magnitude, etc. for each member.

In a cluster environment, 
since most encounters occur at high speed, 
the ones with less-massive galaxies are not expected to have 
significant effects on the evolution of the target galaxy.  
For this reason, given the target as a Milky Way{\textendash}like LTG, 
we apply a mass cut to the Coma members. 
The adopted cut-off is in stellar mass  
$2.1 \times 10^{10}\,h^{-1}{M_{\odot}}$, 
which is about half the total stellar mass of our LTG model~L.  
With the cutoff, 209 galaxies out of 1088 are selected.  
These galaxies are only taken into account for 
the following estimation.

To obtain the 3D distribution of the 209 members,  
we follow the geometrical deprojection method used in \citet{McLaughlin1999}.  
The geometrical technique is simple and fulfills our need 
to estimate an approximate number density profile of the cluster. 
With the assumption of circular symmetry, 
it gives average volume densities 
in a number of concentric spherical shells 
along the 3D deprojected clustercentric radius 
out of the two-dimensional (2D) projected positions 
and the number counts listed in the catalog. 
We describe the procedure for computing the volume densities in Appendix~B.

According to the volume density obtained at each shell, 
we assign the 3D positions of the 209 members by applying 
a Python module generating a random and uniform distribution. 
Because of the random feature we use, 
whenever we generate the 3D distribution of the galaxies,   
the position of the individual one varies 
while satisfying the estimated volume density.  
Thus, to reduce the statistical error, 
we obtain a total of 100 sets for the 3D distribution of 
the 209 galaxies 
and use the median values in our estimation as follows. 
(It should be noted that 
our estimation of the 3D distribution of the galaxies  
is only an approximation because of the small number of galaxies   
and the random feature we used. 
Nevertheless, the estimated information is sufficient for 
our purpose{\textendash-}i.e., as a reference in constructing 
plausible ICs for 
the simulations of the interactions.)

In Figure~3, we show one of the deprojected 3D distribution sets 
in the $x${\textendash}$y$ plane.  
The orbit of a late-type target galaxy is 
chosen as a radial path along the $y$-axis,  
starting from the outer edge $y = 1.9~h^{-1}$~Mpc 
to the cluster center $y = 0$ (blue line). 
Because the orbit is the half of the entire path 
penetrating the cluster radially from one end to another, 
we consider the galaxies located in the upper hemisphere with $y \geq 0$.   
Searching for the galaxies lying relatively close to the path,   
a total of seven galaxies (red circles) are 
found to be located within 70~$h^{-1}$~kpc of the path (shaded region) 
in the upper hemisphere.    
(In the lower hemisphere, four galaxies are found to be 
situated within 70~$h^{-1}$~kpc of the one end of the path near $y = 0$.   
We do not count them.)
The perpendicular distances of these neighboring galaxies to the path 
are 12, 15, 17, 42, 58, 61, and 67~$h^{-1}$~kpc, 
from the closest one to the farthest from the path. 
(Here we consider the completely radial orbit for a simple estimation. 
The LTG path will actually be deflected by the collisions.)

From the entire 100 sets of the distributions, 
we search the closest galaxy from the path in each distribution 
and calculate the median distance of the 100 galaxies. 
We repeat this for the second-closest galaxy, and up to 
the sixth-closest galaxy until  
the median distance does not exceed 70~$h^{-1}$~kpc.
The obtained six median values of the distances 
are 14.6, 24.6, 35.8, 46.3, 53.0, and 61.8~$h^{-1}$kpc, 
from nearest to farthest. 
In other words, an LTG moving radially toward the center  
would likely encounter, on average, 
six galaxies (with at least comparable masses) 
at the closest approach distances 
of roughly 15{\textendash}65~$h^{-1}$~kpc 
at intervals of 10~$h^{-1}$~kpc. 
With the 100 sets, we have checked the tendency that 
more close encounters are likely to occur near the cluster center 
where the galaxy number density is highest, 
and most of the galaxies counted within 70~$h^{-1}$~kpc from the path  
are an early type. 
We have also checked the stellar masses of 
the selected neighboring galaxies from the 100 sets.  
In some distributions, a couple of very massive neighboring galaxies 
with stellar masses $\geq 50 \times 10^{10}\,h^{-1}{M_{\odot}}$ 
are found near the center. 
However, the median values of the stellar masses 
of all selected neighboring galaxies fall in the range 
of 4{\textendash}$5 \times 10^{10}\,h^{-1}{M_{\odot}}$ 
(i.e., about 2{\textendash}2.5 times the stellar mass of our LTG model). 
This justifies the mass ratio of our ETG model to LTG of 2:1.

\begin{figure} [!hbt]
\centering
\includegraphics[width=7.5cm]
{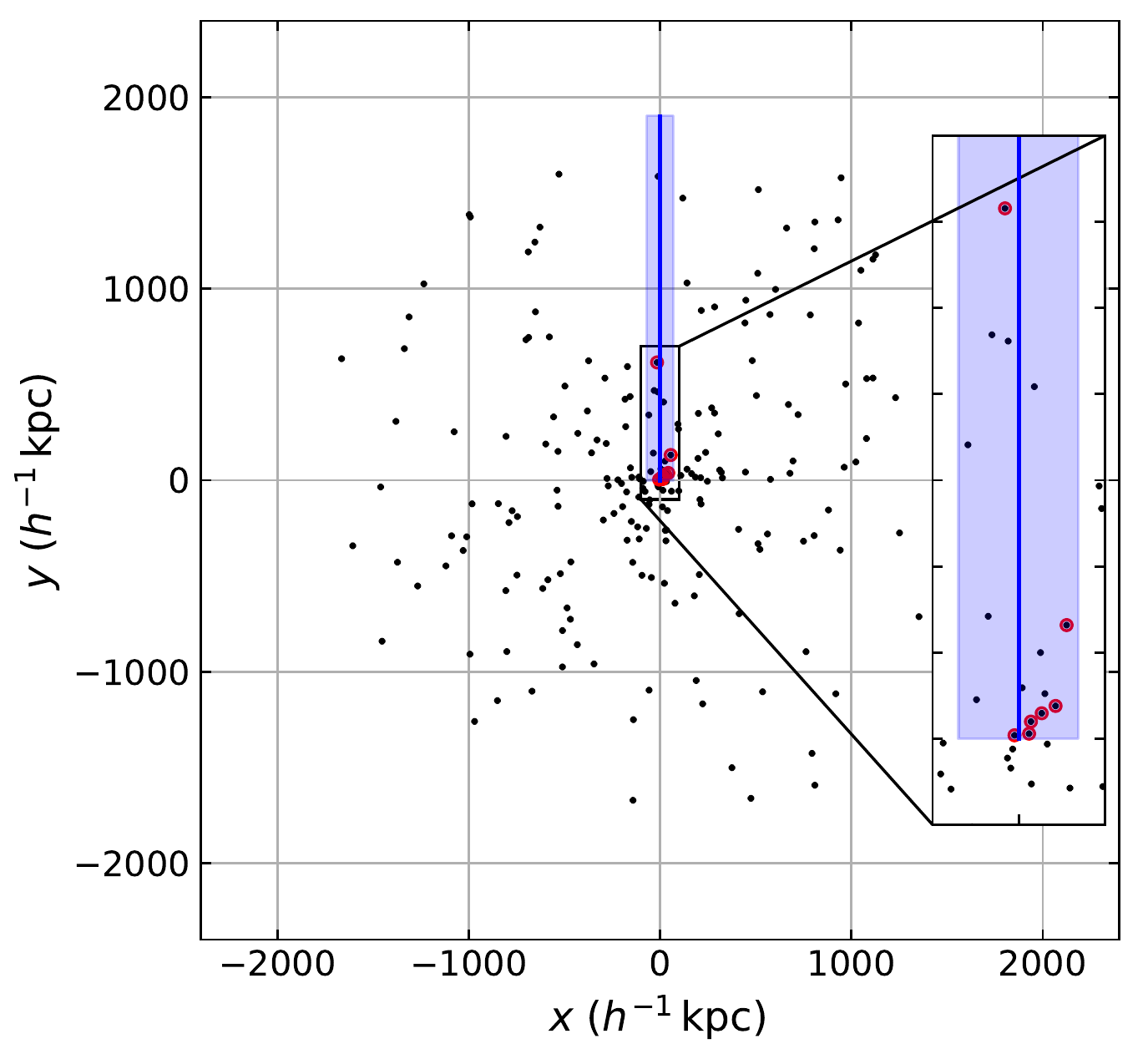}
\caption{One of the deprojected 3D distributions of the 209 
Coma member galaxies (black dots) 
shown in the $x${\textendash}$y$ plane. 
The blue line drawn along the $y$-axis, 
from $y = 1900$ to $y = 0$~$h^{-1}$~kpc ,  
is chosen for the radial path of an LTG galaxy (target galaxy) 
falling toward the cluster center. 
The blue shaded region is the axial cross section of the cylinder with  
a radius of 70~$h^{-1}$~kpc 
whose axis and height coincide with the $y$-axis 
and the orbital path of the target galaxy. 
Within the cylinder, seven galaxies (red circles) are enclosed 
in this distribution (see text for more details). 
The rectangular region $|x|$ $\leq$ $100$~$h^{-1}$~kpc, 
$-100$~$h^{-1}$~kpc $\leq$ $y$ $\leq$ $700$~$h^{-1}$~kpc 
is magnified. }
\end{figure}


\subsection{ICs of the encounter simulations}


\renewcommand{\multirowsetup}{\centering} 
\begin{table*}[htbp]
\caption{Runs of the encounters}
\vspace*{-2mm}
\begin{center}
\begin{tabular}{l|c c|c c | c c c c c}
\hline \hline
\multirow{3}{0.8cm}{Runs\footnotemark[1]}& \multicolumn{2}{p{3.3cm}|}{\centering LTG Model}  & \multicolumn{2}{p{3.3cm}|}{\centering ETG Model\footnotemark[2]} & 
\multicolumn{3}{p{4.0cm}}{\centering At the Closest Approach} \\
& \multicolumn{1}{c}{Initial $x$, $y$, $z$} 
& \multicolumn{1}{c|}{Initial $v_x$, $v_y$, $v_z$}  
& \multicolumn{1}{c}{Initial $x$, $y$, $z$} 
& \multicolumn{1}{c|}{Initial $v_x$, $v_y$, $v_z$} 
& \multicolumn{1}{c}{$d_{i}\footnotemark[3]$}  
& \multicolumn{1}{c}{$dv_{i}\footnotemark[4]$} 
& \multicolumn{1}{c}{$t_i\footnotemark[5]$}  
& \multicolumn{1}{c}{$t_{{\rm gg}}\footnotemark[6]$} 
& \multicolumn{1}{c}{$t_{\rm g}\footnotemark[7]$}
\\ 
& \multicolumn{1}{c}{($h^{-1}$kpc)} 
& \multicolumn{1}{c|}{(km~s$^{-1}$)} 
& \multicolumn{1}{c} {($h^{-1}$kpc)} 
& \multicolumn{1}{c|}{(km~s$^{-1}$)} 
& \multicolumn{1}{c} {($h^{-1}$kpc)} 
& \multicolumn{1}{c} {(km~s$^{-1}$)} 
& \multicolumn{1}{c} {($h^{-1}$Gyr)} 
& \multicolumn{1}{c} {($h^{-1}$Gyr)} 
& \multicolumn{1}{c} {($h^{-1}$Gyr)}   
\\      
\hline
1e, 1f & $-543$, 68, 0 & 1500, 0, 0  &    0,  0, 0 & 0, 0, 0 & $d_{1} = 65$ & $dv_{1}$ = 1569 & $t_{1} = 0.35$ & 0.35 & 1.05\\
2e, 2f &    104, 67, 0 & 1526, 0, 0  &  649,  7, 0 & 0, 0, 0 & $d_{2} = 55$ & $dv_{2}$ = 1583 & $t_{2} = 0.35$ & 0.77 & 1.47 \\
3e, 3f &    753, 64, 0 & 1519, 0, 0  & 1296, 18, 0 & 0, 0, 0 & $d_{3} = 45$ & $dv_{3}$ = 1583 & $t_{3} = 0.35$ & 1.19 & 1.89 \\
4e, 4f &   1340, 65, 0 & 1514, 0, 0  & 1940, 29, 0 & 0, 0, 0 & $d_{4} = 35$ & $dv_{4}$ = 1586 & $t_{4} = 0.35$ & 1.61 & 2.31 \\
5e, 5f &   2042, 65, 0 & 1500, 0, 0  & 2579, 42, 0 & 0, 0, 0 & $d_{5} = 25$ & $dv_{5}$ = 1591 & $t_{5} = 0.35$ & 2.03 & 2.73 \\
6e, 6f &   2681, 69, 0 & 1490, 0, 0  & 3213, 59, 0 & 0, 0, 0 & $d_{6} = 15$ & $dv_{6}$ = 1608 & $t_{6} = 0.35$ & 2.45 & 3.15 \\   
\hline
\end{tabular}
\footnotetext[1]{The run named ``$i$e" or ``$i$f" ($i$ = 1, 2, ..., 6) 
represents the $i$th edge-on or face-on  
encounter simulation between models~L and EH$_i$.}
\footnotetext[2]{In the $i$th encounter runs ($i$ = 1, 2, ..., 6), 
the initial positions and velocities of the ETG model 
mean those of model~EH$_i$. 
These initial values of model~EH$_i$ (and of model~L) 
at the start of the $i$th encounter runs are the same 
for both edge-on and face-on cases.}
\footnotetext[3]{$d_{i}$ ($i$ = 1, 2, ..., 6) represents 
the distance between models~L and EH$_i$  
in the $i$th encounter simulations. 
The values of $d_{i}$ for the edge-on and the face-on cases 
in the given $i$th encounter runs    
are not always exactly the same but almost equal.  
When the values are not exactly matched to each other, 
we list the values from the edge-on case throughout this paper.}
\footnotetext[4]{$dv_{i}$ ($i$ = 1, 2, ..., 6) is 
the relative velocity between models~L and EH$_i$ 
in the $i$th encounter simulations. 
As in $d_{i}$, the values of $dv_i$ for the edge-on 
and face-on cases are not always exactly equal.} 
\footnotetext[5]{$t_{i}$ ($i$ = 1, 2, ..., 6) is the time measured since 
the start of the $i$th encounter runs. 
The $i$th collisions (closest approaches) both in the edge-on 
and the face-on cases occur at $t_{i} = 0.35$~$h^{-1}$Gyr.}
\footnotetext[6]{$t_{\rm gg}$ is the accumulated time elapsed    
since the start of the first-encounter runs, i.e., from $t_1 = 0$.} 
\footnotetext[7]{$t_{\rm g}$ is the accumulated time elapsed 
since the start of the runs with each of 
models~L and EH$_1$ in isolation 
(i.e., $t_{\rm g}$ = $t_{\rm gg} + 0.7$~$h^{-1}$Gyr). 
} 
\end{center}
\end{table*}

\begin{figure*} [!hbt]
\centering
\includegraphics[width=15.0cm]{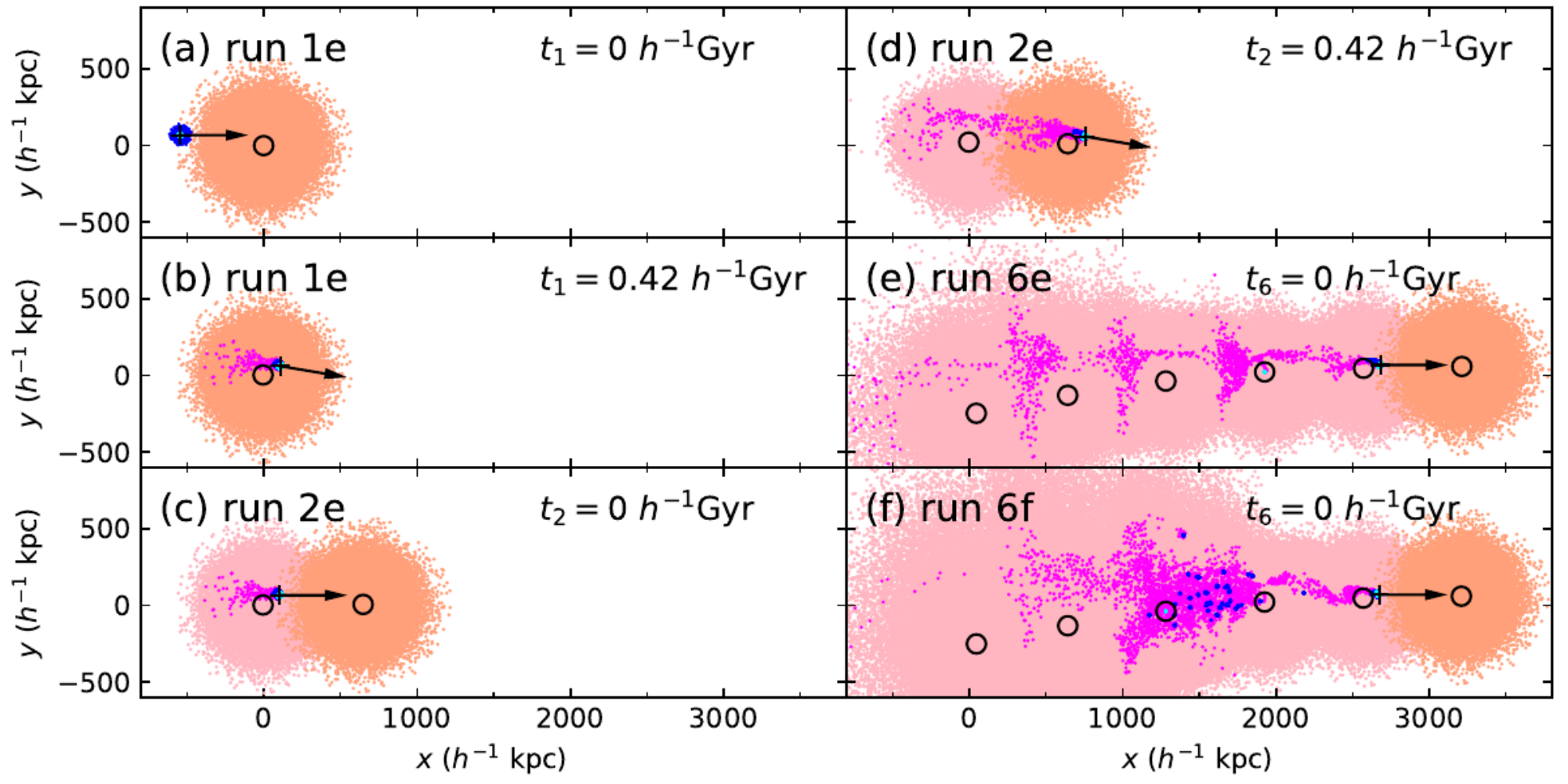}
\caption{
Distribution of the gas particles of the LTG and ETG models 
from the encounter simulations seen in the $x${\textendash}$y$ plane. 
The positions of the center of the LTG and each ETG model are marked 
with a plus and a circle, respectively. 
The direction of the motion of the LTG model~L is indicated by 
an arrow.       
The different colors of the gas particles 
are used in the same way as in Figure~2.    
(a) Initial configuration of run~1e at $t_1$ = 0. 
Model~L has an initial velocity of 1500~km~s$^{-1}$ in the 
positive $x$-axis (refer to Table~2).
(b) Snapshot of run~1e at $t_1$ = 0.42~$h^{-1}$~Gyr, 
which is 0.07~$h^{-1}$~Gyr after the closest approach 
between models~L and EH$_{1}$. 
The velocity of model~L is ($v_x$, $v_y$, $v_z$) = (1524, $-$78, 0)~km~s$^{-1}$, 
and the direction is facing $\sim$3$^\circ$ below the $x$-axis 
(the direction and the length of the arrow are not scaled). 
Model~EH$_{1}$ is moving slowly 
with the velocity of ($v_x$, $v_y$, $v_z$) = ($-$12, 42, 0)~km~s$^{-1}$  
(the direction of the motion of model~EH$_{1}$ is not shown). 
(c) Initial configuration of run~2e at $t_2$ = 0. 
The snapshot of the above panel(b) is rotated $\sim$3$^\circ$ 
around tje $z$-axis in a counterclockwise direction, and then 
model~EH$_{2}$ is included at 
($x$, $y$, $z$) = (649, 7, 0)~$h^{-1}$~kpc in the rotated coordinate. 
The direction of motion of model~L is now completely horizontal. 
(d) Snapshot of run~2e at $t_2$ = 0.42~$h^{-1}$~Gyr,  
which is 0.07~$h^{-1}$~Gyr after the closest approach 
between models~L and EH$_{2}$. 
The direction of motion of Model~L is 
slightly below the $x$-axis. 
(e){\textendash}(f) Initial configuration of run~6e and 6f, respectively, 
at $t_6$ = 0. The direction of motion of model~L is horizontal. 
Model~EH$_{6}$ is included at ($x$, $y$, $z$) = (3213, 59, 0)~$h^{-1}$~kpc . 
}
\end{figure*}

Based on the estimation using the 
Coma cluster members (explained in Section~4.1), 
we design the multiple encounters of 
an LTG with neighboring ETGs 
using our galaxy models~L and EH (Table~2). 
For a radial orbit from the outskirts 
to the cluster center, 
the late-type target galaxy is 
intended to collide consecutively with six ETGs, 
first at the closest approach distances of 
65~$h^{-1}$~kpc ($d_{1}$) 
and then at 55 ($d_{2}$), 45 ($d_{3}$), 35 ($d_{4}$), 25 ($d_{5}$), 
and 15~$h^{-1}$~kpc ($d_{6}$) in order.  
(Hereafter, we will use the subscripts 1 through 6 to 
distinguish the distances between the two models
(and other quantities)  
of the first through sixth encounters.  
For all of the colliding ETGs, the same model~EH 
is used six times and called 
model~EH$_1$ through model~EH$_6$.)  
The consecutive encounters are considered 
for the two different disk tilt angles 
relative to the orbital direction of the LTG, 
either edge-on or face-on, keeping all other parameters fixed.  
Since the ETGs would not affect the LTG significantly 
at great distances, the ETGs are included one by one 
in our ICs of the encounter simulations 
when the LTG approaches each of them relatively closely as follows.

The ICs of the first-encounter simulations for both 
edge-on and face-on cases (runs~1e and 1f, respectively)  
are presented in Table~2 and Figure~4(a).  
Model~EH$_1$ is positioned at the origin, and 
model~L is placed far enough away from the ETG 
at ($x_{0}$, $y_{0}$, $z_{0}$) = ($-$543, 68, 0)~$h^{-1}$~kpc, 
about three times the virial radius of model~EH$_{1}$ apart. 
At the initial time, model~EH$_{1}$ is stationary 
and model~L has a velocity of 
($v_{x0}$, $v_{y0}$, $v_{z0}$) = (1500, 0, 0)~km~s$^{-1}$ 
in the horizontal direction toward the ETG. 
Both LTG and ETG models here are included 
after 0.7~$h^{-1}$~Gyr evolution in isolation.    
The initial position of model~L relative to model~EH$_{1}$ is chosen 
so that the two galaxy models encounter most closely 
with $d_1 = 65$~$h^{-1}$~kpc at $t_1 = 0.35~h^{-1}$~Gyr 
($t_1$ represents the time 
elapsed since the start of the first-encounter simulations).     
Figure~4(b) shows the snapshot of run~1e 
at $t_1$ = 0.42~$h^{-1}$~Gyr, 
shortly after the collision between models~L and EH$_{1}$.   
The position ($x$, $y$, $z$) and the velocity ($v_x$, $v_y$, $v_z$) 
of model~L at this time are 
(108, 61, 0)~$h^{-1}$~kpc and 
(1524, $-$78, 0)~km~s$^{-1}$, respectively. 
Those of model~EH$_{1}$ are 
($-$4, 4, 0)~$h^{-1}$~kpc and 
($-$12, 42, 0)~km~s$^{-1}$. 
The separation between the two models is now 
126~$h^{-1}$~kpc, and  
the direction of motion of model~L 
is slightly ($\sim$3$^\circ$) below the horizontal axis. 
The second ETG is about to be included at this time step, 
as explained below.

The ICs of runs~2e and 2f  
are created by using the snapshots taken at $t_1$ = 0.42~$h^{-1}$~Gyr 
from runs~1e and 1f, respectively.  
Before putting the second ETG model~EH$_{2}$ 
(which is identical to model~EH$_{1}$ 
and has evolved for 0.7~$h^{-1}$~Gyr in isolation as well), 
the snapshots are rotated until  
the direction of motion of model~L becomes completely horizontal 
(i.e., $\sim$3$^\circ$ around the $z$-axis in the 
counterclockwise direction).
Model~EH$_{2}$ is then added at the right side of model~L, 
as shown in Figure~4(c) 
at (649, 7, 0)$~h^{-1}$~kpc in the rotated coordinate.   
(The rotation does not alter the relative positions of the models. 
It is done solely for convenience in showing the deflection of model~L 
during each encounter simulation.)     
The position of model~EH$_{2}$ is chosen  
so that the closest approach between models~L and EH$_2$ 
occurs at $t_2 = 0.35~h^{-1}$~Gyr 
with $d_2 = 55$~$h^{-1}$~kpc. 
The snapshot taken at $t_2$ = 0.42~$h^{-1}$~Gyr, 
shortly after the second collision, is displayed in Figure~4(d).  
The motion of model~L is heading slightly downward 
below the $x$-axis. 
The snapshot is rotated again until the direction of model~L 
becomes completely horizontal, and then 
the third ETG is added at (1296, 18, 0)~$h^{-1}$~kpc   
for the simulations of the third encounter. 
The above steps are repeated until the sixth ETG is included 
in the snapshot at $t_5$ = 0.42~$h^{-1}$Gyr. 
Figure~4(e) and (f) present the configuration of the LTG and 
the six ETGs 
at the start of the sixth-encounter simulations 
for the edge-on and face-on cases, respectively.


\section{Evolution of an LTG in a cluster environment}

As noted earlier, 
an LTG falling into a cluster will interact with 
nearby cluster member galaxies 
and the cluster at the same time. 
However, taking both 
galaxy{\textendash}galaxy and galaxy{\textendash}cluster interactions 
into account simultaneously 
in the hydrodynamic simulations 
is complicated and overly time-consuming 
because of the very different scales/properties of 
the galactic and cluster 
components (i.e., size, mass, density, temperature, etc).
Our simulations focus on galaxy{\textendash}galaxy interactions 
with the aim of examining how and to what extent 
the multiple galaxy interactions  
can affect the evolution of cluster LTGs. 
We then compare our simulation results with 
those of galaxy{\textendash}cluster interactions. 
Here we first present our simulation results  
and then the comparison with others.


\subsection{Results of our simulations: effects of multiple galaxy{\textendash}galaxy interactions}

\begin{figure*} [!hbt]
\centering
\includegraphics[width=16.5cm]{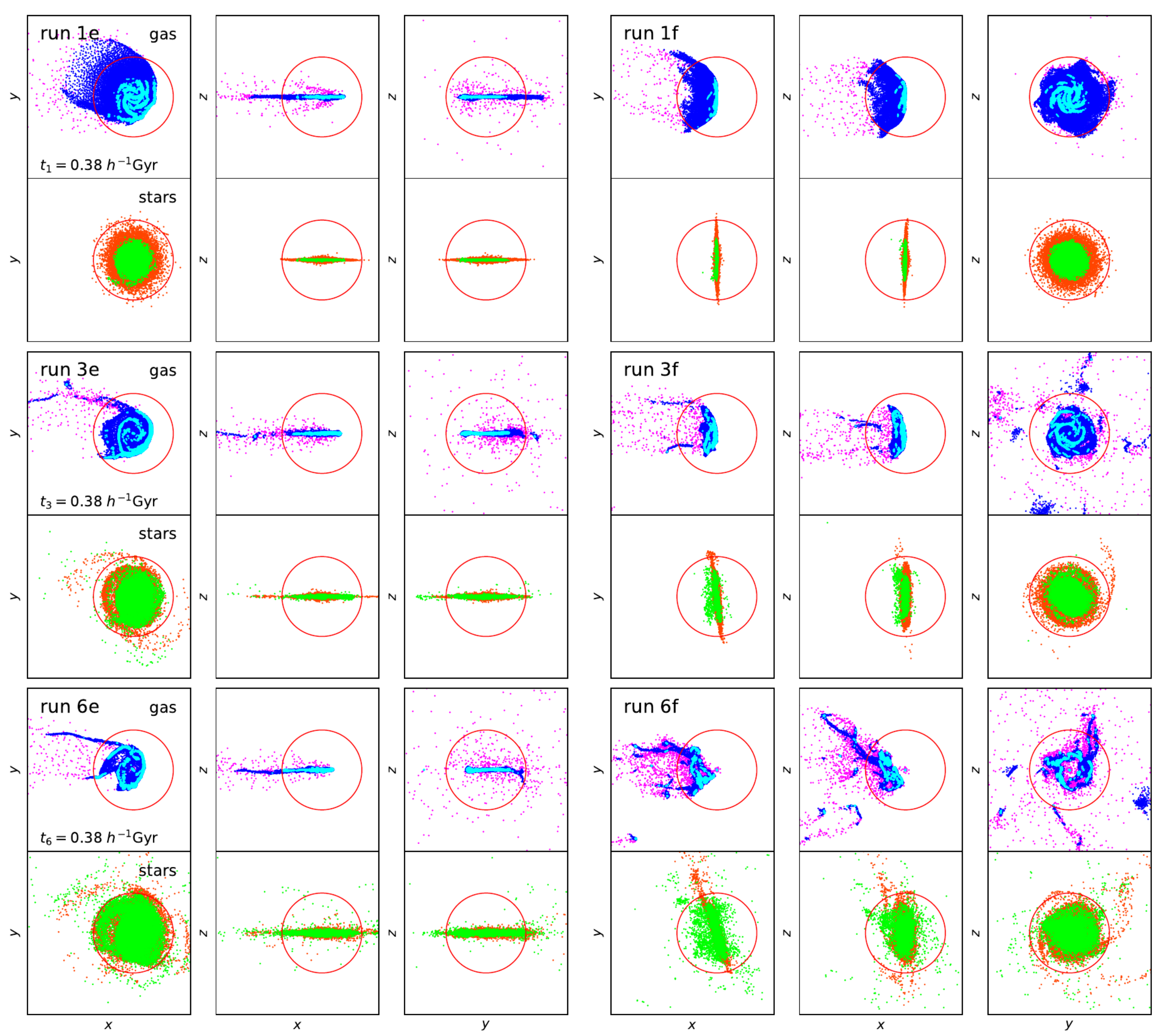}
\caption{
Distribution of the disk particles of model~L 
at $t_1$ = 0.38~$h^{-1}$~Gyr (top two rows), 
$t_3$ = 0.38~$h^{-1}$~Gyr (middle two rows), 
and $t_6$ = 0.38~$h^{-1}$~Gyr (bottom two rows) 
in the edge-on (left three columns) and 
face-on (right three columns) cases. 
The gas and star particles are presented separately 
(upper and lower panels, respectively) 
in orthogonal projections.   
The disk gas particles are displayed with different colors 
the same way as in Figure~4, i.e., 
hot non-star-forming gas in magenta, 
cold non-star-forming gas in blue, 
and star-forming gas in cyan. 
The stars originally set as the disk star particles of model~L are 
plotted in orange, 
and the stars added onto the disk out of the initial disk gas 
are plotted in green. 
Each panel measures 100~$h^{-1}$~kpc 
on a side. 
A sphere is drawn (red) 
in the center of the disk 
with a radius of 24.5~$h^{-1}$~kpc,   
which enclosed 90\% of the disk gas 
at the initial time $t_1$ = 0. 
}
\end{figure*}

\begin{figure} [!hbt]
\centering
\includegraphics[width=8.3cm]{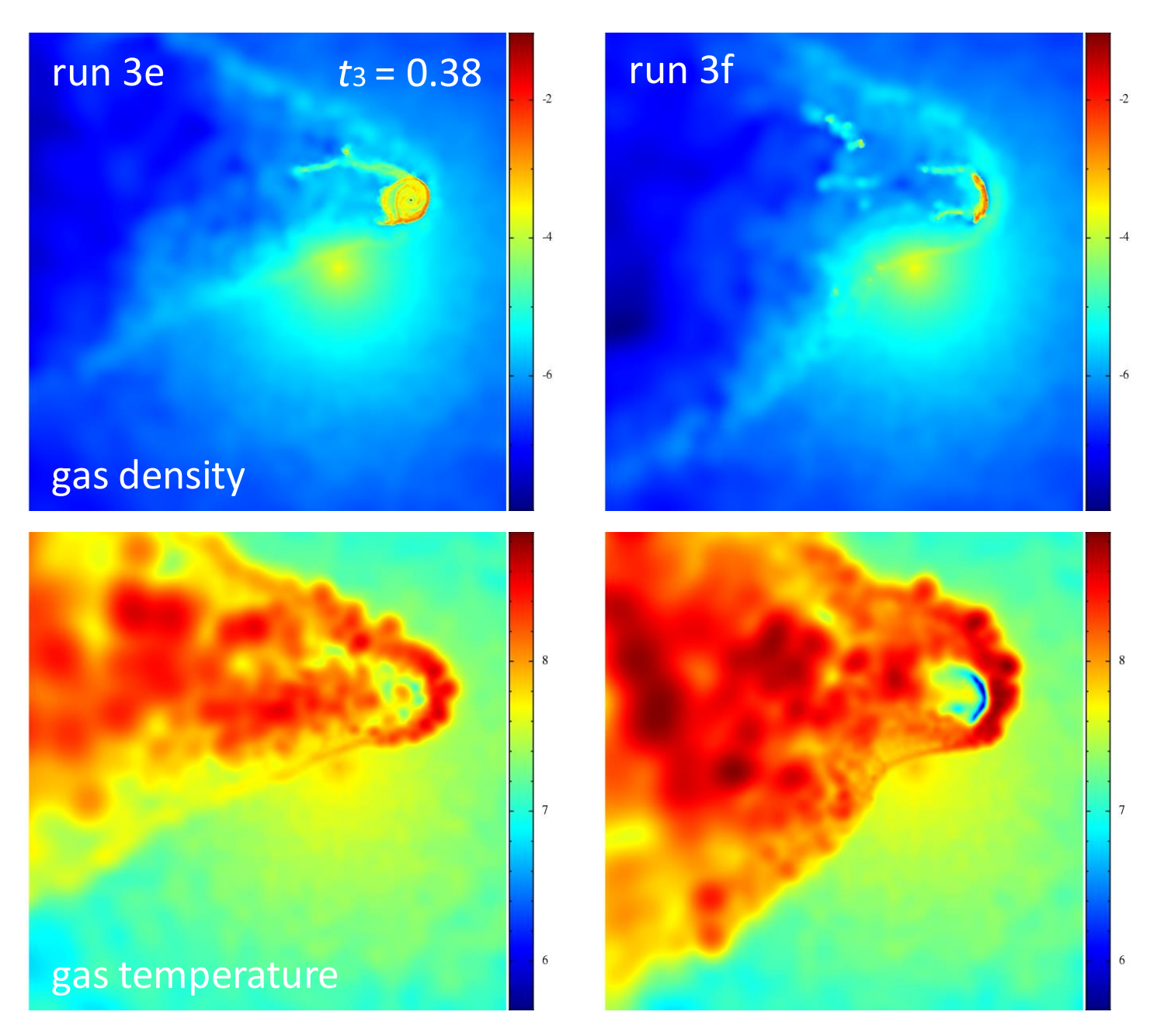}
\caption{Log column density (top row) and temperature (bottom row) 
of the gas components from runs~3e (left column) and 3f (right column) 
at $t_3$ = 0.38~$h^{-1}$~Gyr, 
projected in the $x${\textendash}$y$ plane. 
Only the gas particles within $\mid{z}\mid\,\leq 10$ are taken into account. 
Each of the color ranges for the density and temperature are fixed 
for both edge-on and face-on cases.  
Each panel measures 300~$h^{-1}$~kpc on a side.  
This figure is made using SPLASH (\citealt{Price2007}).  
}
\end{figure}

\begin{figure*} [!hbt]
\centering
\includegraphics[width=16.5cm]{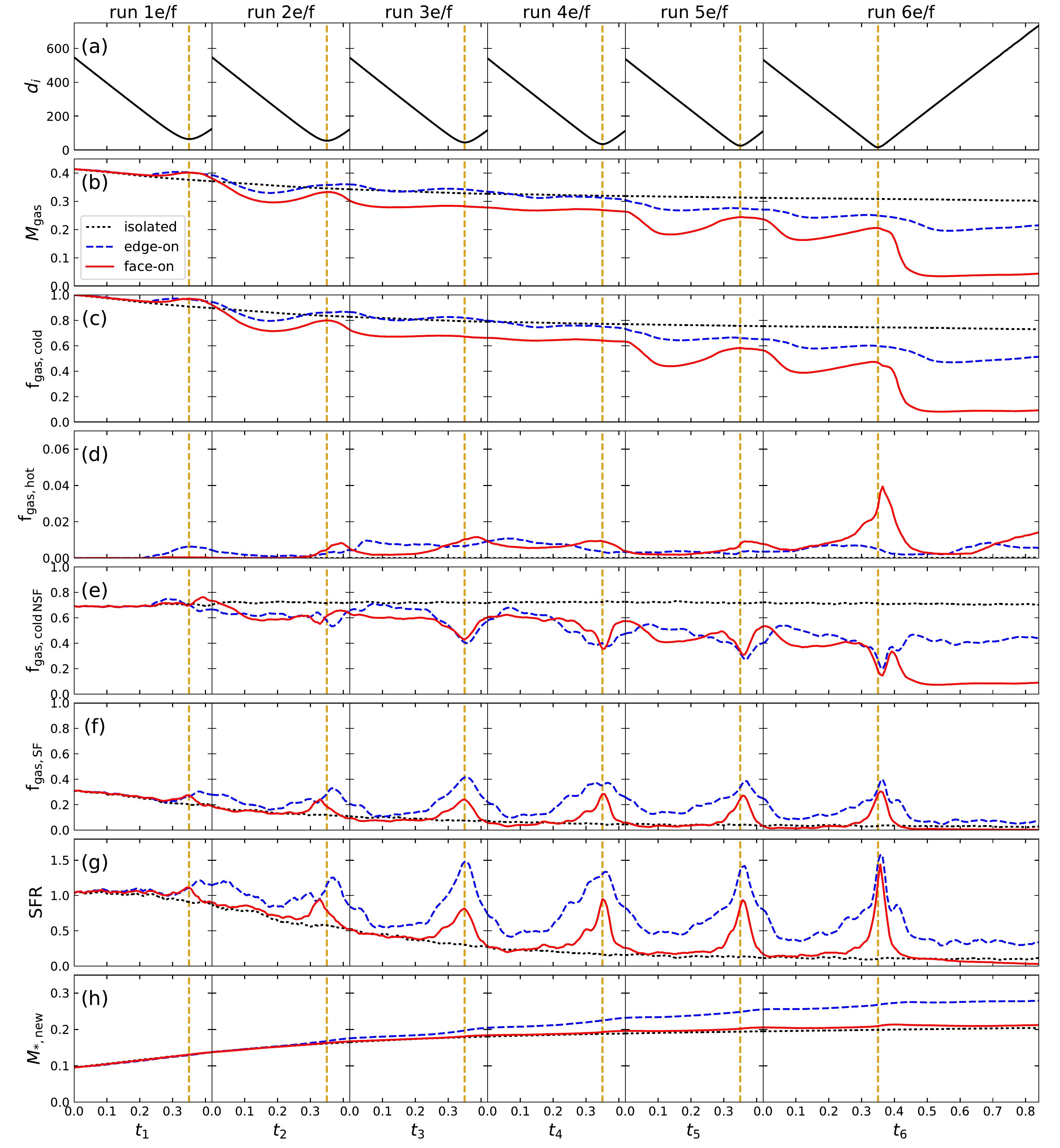}
\caption{
Variation of some disk quantities throughout the simulations. 
From left to the right columns, 
the values obtained from the first-encounter 
simulations (runs 1e and 1f) through 
the sixth-encounter simulations (runs~6e and 6f) are displayed. 
The units are $h^{-1}$~Gyr for time, 
$h^{-1}$~kpc for distance, 
$10^{10}\,h^{-1}\,{M_{\odot}}$ for mass, 
and ${M_{\odot}}$~yr$^{-1}$ for SFR. 
The vertical dashed line drawn at $t_{i}$ = 0.35~$h^{-1}$~Gyr 
($i$ = 1, 2, ..., 6) 
in each column indicates 
the time of the closest approach between 
models~L and EH$_{i}$.     
(a) Distance between model~L and model~EH$_{i}$ ($i$ = 1, 2, ..., 6) 
in the $i$th encounter runs. 
The values from the edge-on and face-on cases are 
almost equal to each other. 
(b){\textendash}(h) Quantities of the disk particles 
of model~L enclosed within the sphere shown in Figure~5. 
The dotted, dashed, and solid lines 
represent the quantities from the isolated, edge-on, 
and face-on cases, respectively. 
(b) $M_{\rm gas}$ = total mass of the gas particles (i.e., blue, cyan, and magenta particles 
enclosed within the sphere in Figure~5). 
(c) $f_{\rm gas,\,cold}$ = total mass of the cold gas (blue and cyan) / 
total mass of the gas at $t_1$ = 0. 
(d) $f_{\rm gas,\,hot}$ = total mass of the hot gas (magenta) / 
total mass of the gas at $t_1$ = 0. 
(e) $f_{\rm gas,\,cold\,NSF}$ = total mass of the cold non-star-forming gas (blue) / 
total mass of the gas at $t_1$ = 0. 
(f) $f_{\rm gas,\,SF}$ = total mass of the star-forming gas (cyan) / 
total mass of the gas at $t_1$ = 0. 
(g) SFR = sum of the SFRs of the star-forming gas (cyan). 
(h) $M_{*,\rm{new}}$ = total mass of the new star particles formed out of the gas 
(i.e., green particles enclosed within the sphere in Figure 5). 
}
\end{figure*}

We begin with Figure~2, which shows  
the evolution of 
the temperature ($T$) and 
local density ($\rho_{\rm loc}$) for the gas components 
of the LTG and ETG models  
in both edge-on and face-on encounter simulations. 
The initial values of $\rho_{\rm loc}$ and $T$, 
at the start of the first-encounter simulations of runs~1e and 1f, 
are presented in the two top right panels of the figure. 
At the initial time of $t_1 = 0$,    
the disk gas particles of model~L (left panel) are situated 
lower right side in the $\rho_{\rm loc}${\textendash}$T$ plane. 
The star-forming gas (plotted in cyan) 
follows the fingerlike pattern in the higher-density region, 
which results from the effective equation of state for star-forming gas
in the multiphase model (\citealt{Springel_Hernquist2003}).
The halo gas particles of model~EH$_{1}$ (right panel) 
are located upper left side in the plane at the initial time. 
The dashed line is used as a fixed criterion to divide 
``cold" gas particles (those lying below the line) 
from ``hot" gas particles (those lying above the line) 
throughout the simulations. 
At $t_1 = 0$, when the two galaxies have not yet started interactions, 
the disk gas and halo gas are well separated by the dashed line.

The $\rho_{\rm loc}${\textendash}$T$ distribution of the gas 
shortly after model~L collided with models~EH$_{1}$, EH$_{3}$, 
and EH$_{6}$ is shown in the second through fourth rows of Figure~2. 
Some of the particles initially set as the disk gas of model~L  
are found above the dashed line (magenta dots) in the 
relatively high-temperature and low-density region, 
as they are heated and stripped off the disk 
through interactions with the ETGs.  
This hot gas appears more as the LTG experiences more collisions and 
when it flies face-on rather than edge-on.

The snapshots showing the appearance of 
the gas and the stellar disks 
of model~L 
at the same times as in Figure~2 are 
presented in Figure~5. 
The gas disk forms the bow-like front 
as it moves fast against the halo gas included in the ETGs (cf. Figure~4). 
The star-forming gas particles (cyan) 
are found mostly  
along the spiral arms and the 
shock front. 
They will subsequently turn into stars (green) according to 
the star formation rates (SFRs) of the gas particles.   
Some of the disk gas is heated (magenta) and stripped off the disk.  
The long gas tail developed in runs~3e and 6e 
(see the $x${\textendash}$y$ views) 
is the result of the shock boundary combined with 
the clockwise directional rotation of the disk. 
The corresponding stellar disk, in contrast, forms  
more symmetrical two-sided tails.  
Overall, the gas disk shows very different morphology 
compared with the stellar one,  
due to hydrodynamic interactions 
with the halo gas of the ETGs.  
The offset between the ``new" stars (green; stars formed out of the disk gas) 
and ``old" stars (orange; stars originally set as the disk stars)  
is caused by the shock. 
Figure~6 more clearly shows the shock that arises  
from the collision between the gas disk of model~L 
and the gas halo of model~EH$_3$ at $t_3$ = 0.38~$h^{-1}$~Gyr,  
shortly after the third encounter. 
The shock developed in the face-on case is wider 
than that in the edge-on case.

In order to examine the evolution of the disk materials, 
we compute various quantities of the disk particles  
enclosed within a fixed sphere around the center of the disks 
over time and present the results in Figure~7.  
The large circle (red) shown in Figure~5   
is the cross section of the sphere with a radius of 4~$\times$~$a_{dg}$,  
which initially enclosed 90$\%$ of the total mass of the disk gas 
at $t_{1} = 0$. 
As displayed in Figure~7(b), 
the LTG loses more disk gas (which is initially set as the disk gas) 
through the collisions 
compared with the isolated disk (black dotted line). 
The disk gas, which is comprised of (cold) star-forming gas, 
cold non-star-forming gas, and hot (non-star-forming) gas (cf. Figure~2), 
decreases more severely in the face-on case (red solid line) 
than in the edge-on case (blue dashed line), 
in particular, most abruptly after the sixth, deepest collision.   
The cold gas (star-forming $+$ non-star-forming) fraction  
also drops shortly after each collision (panel (c)). 
In the face-on case, the remaining cold gas within the sphere 
after the sixth collision 
is only $\sim$10\% of the initial value. 
The hot gas fraction in the edge-on and face-on cases becomes 
positive due to the collisions, whereas it remains zero 
in the isolated case (panel (d)).
The hot gas fraction rises dramatically near the sixth face-on collision. 
Among the cold gas, the fraction 
of non-star-forming gas and 
star-forming gas decreases and increases in the opposite way 
near each collision (panels (e) and (f), respectively). 
The edge-on collisions produce star-forming gas more efficiently 
than the face-on collisions by 
inducing successive compression on the disk.  
By the same token, the SFR in the edge-on case is overall greater 
than in the face-on case, as well as in the isolated case (panel (g)). 
The SFR in the face-on case increases only near the collisions and 
is otherwise similar to that in the isolated case; 
it finally becomes lower than that in the isolated case 
after the sixth collision, 
when the disk loses a significant amount of cold gas. 
More stars are added onto the disk out of the star-forming gas in the 
edge-on case than in the face-on and isolated cases (panel (h)).


\subsection{Comparison to other simulations: Effects of galaxy{\textendash}cluster interactions}

Here we refer to the numerical study of \citet{Jachym+2007} 
to compare with ours.  
The reason we choose 
this work among many others is    
because they considered the interactions between 
an LTG model comparable to ours and a cluster model 
along a completely radial orbit, just like in our work.  
They also used the $N$-body/SPH code GADGET 
(version 1.1; \citealp{Springel+2001}) for the simulations, 
with some modifications as described below, 
which allows us more direct comparison.

In \citet{Jachym+2007}, 
the standard galaxy model (``LM" in the paper) 
is a Milky Way{\textendash}like model 
consisting of both stellar and gas disks, a stellar bulge, 
and a DM halo. 
(We use their results obtained with only the standard galaxy model~LM.)   
The total disk mass and disk gas fraction are    
$8.6 \times 10^{10}{M_{\odot}}$ and 0.1 in mass, respectively.  
Their standard cluster model is 
a Virgo cluster{\textendash}like model possessing both DM and gaseous halos. 
The gas component of the hot ICM  
follows a $\beta$-profile (Equation~(4) of the paper) 
with $R_{\rm{c, ICM}}$ = 13.4~kpc (a parameter of the ICM central concentration) 
and $\rho_{0, \rm{ICM}}$ = $4 \times 10^{-3}\rm{cm^{-3}}$ 
(the volume density of the ICM in the cluster center). 
To model a wide variety of clusters 
from rich ones having a lot of hot gas 
to poor ones with only a little gas, 
they vary the values of $R_{\rm{c,ICM}}$ and $\rho_{0, \rm{ICM}}$ 
from 4 times to a quarter of the standard values, respectively.

They made some modifications in the GADGET code 
in order to simulate the hydrodynamic interactions 
between the disk gas and the cluster gas more appropriately.   
Whereas the original code handles all of the gas particles in one group,  
the modified code treats the disk and cluster gas 
as two different groups 
with different spatial resolution  
(for details, refer to Section~5 of their paper). 
This reduces numerical artifacts that can occur when the 
number densities of the two gas components are significantly different.  
The numbers of disk gas and cluster gas particles 
used in the standard models are 12,000 and 120,000, respectively. 
To achieve a comparable number density for the 
cluster gas component to that of the disk gas component 
using the limited number of particles, 
they kept all of the cluster gas within the 140~kpc radius, 
applying periodic boundary conditions. 
(It is not desirable, but rather inevitable 
to save computational expenses.)

In their simulations,   
the LTG model moves face-on 
in a completely radial orbit, 
from one edge of the cluster model 
passing the center to the other. 
The distance from the outskirts of the cluster to the center is 1~Mpc. 
The times when the LTG 
enters the ICM region, passes the center, 
and escapes the ICM region are 
about 1.52, 1.64, and 1.76~Gyr, respectively, 
since the start of the run. 
The velocity when it passes the center of the cluster 
is about 1300~km~s$^{-1}$.
In the run passing through the standard cluster model, 
the LTG model turned out to lose   
about one-third of its original disk gas 
by the end of the run at 2~Gyr.  
More specifically, 
the LTG model continues to lose its gas 
until just after it passes the cluster center. 
The minimum mass of the gas disk ($M_{\rm min}$),  
measuring the gas within $\pm 1$~kpc from the midplane of the disk,  
is 49~\% of the original mass 
about 20~Myr after it passes the center. 
After the minimum, much of the stripped gas  
become accreted back onto the disk. 
The mass of the reaccreted gas ($M_{\rm accr}$) is 22~\% of the 
original disk mass. 
Thus, the mass of the stripped gas ($M_{\rm strip}$) is 29~\% 
of the initial mass of the gas disk 
(cf. $M_{\rm min}$ + $M_{\rm accr}$ + $M_{\rm strip}$ = 100~\%).

In the run with the cluster model containing the richest ICM 
(having four times greater values of $R_{\rm{c, ICM}}$ and 
$\rho_{0, \rm{ICM}}$ 
than those of the standard cluster model), 
$M_{\rm min}$,  $M_{\rm accr}$, and $M_{\rm strip}$ are 
15~\%, 0~\%, and 85~\%, respectively. 
In the opposite case, with the poorest ICM in the cluster model, 
$M_{\rm min}$, $M_{\rm accr}$, and $M_{\rm strip}$ are 
84~\%, 15~\%, and 1~\%, respectively.

Considering the fact that 
their standard cluster model 
is designed to represent a Virgo-like cluster,  
the ICM is confined within the sphere of the radius 140~kpc, 
and the maximum relative velocity 
of the LTG model is about 1300~km~s$^{-1}$ 
(which is about 300~km~s$^{-1}$ lower than that of ours), 
the standard cluster run may not be equivalent 
to make a comparison with our results,  
although their LTG orbit is a full path 
from one edge of the cluster model to the other.

Instead, their simulation using the cluster model with the richest ICM 
would be more suitable for the comparison. 
In both the richest ICM run  
and our face-on run, 
the LTGs came out to lose most of the gas 
through interactions with either the cluster 
or the six neighboring ETGs with hot gas, respectively.  
The amounts of the stripped gas obtained from the two runs 
are almost equivalent to each other, 
as both fall in the range of 80\%{\textendash}90\% 
of the initial gas. 
This implies that the impact  
on disk gas stripping by hydrodynamic interactions 
with the cluster gas or the hot gas of 
many neighboring galaxies could be equally significant. 
Of course, because the distributions of both ICM and 
member galaxies in a cluster 
are not uniform, the LTGs in a cluster could evolve  
very differently, depending on the dynamical histories.


\section{Summary and discussion}

We have examined the evolution of LTGs 
in a cluster environment, focusing on the 
effects of high-speed multiple interactions with cluster ETGs,  
using $N$-body/SPH simulations. 
For this, we built the LTG model~L 
having a gas disk 
and the ETG model~EH containing a gas halo, 
with a total mass ratio of the LTG to ETG of 1:2. 
Based on the deprojected distribution of the Coma cluster members, 
we set the ICs of the consecutive collisions of an LTG 
with six ETGs, at the closest approach distances of 
65, 55, 45, 35, 25, and 15~$h^{-1}~kpc$ in order, 
at the relative velocities of about 1500{\textendash}1600~km~$s^{-1}$  
for either edge-on or face-on motion of the LTG.

We find that the evolution of an LTG can be significantly affected 
by high-speed multiple collisions with ETGs, particularly 
through the hydrodynamic interactions 
between the cold disk of the LTG 
and the hot gas halos of the ETGs. 
The LTG model~L loses about half of its initial cold gas 
after the six collisions in edge-on, while  
the isolated disk loses about 25~\% 
during the same period (Figure~7 (c)). 
For the face-on collision case, 
the cold gas removal from model~L during the same period 
reaches about 90~\% of its original gas 
due to the strongest ram pressure exerted on the disk.   
The amount of stripped gas obtained from our face-on consecutive run 
is as much as that found in the simulations of galaxy{\textendash}cluster interactions of \citet{Jachym+2007}, 
where an LTG (which is comparable with ours)  
is flying through a cluster with rich ICM (Section~5.2). 
This means that the evolution of LTGs in a cluster 
can be strongly affected by the interactions with not only 
the cluster but also the neighboring galaxies possessing hot gas.  
Depending on the dynamical history, the LTGs could be more influenced   
by collisions with neighboring member galaxies.

Our simulations show that star formation is enhanced 
in the LTG through consecutive high-speed collisions 
with the ETGs containing hot gas (Figure~7 (g)). 
For the LTG flying edge-on, not only does the SFR rise near the collisions,   
it also remains higher at all times than that in the LTG flying face-on. 
The pressure exerted on the leading side of the edge-on disk 
by collisions with hot halos subsequently compresses the 
cold disk more efficiently, 
leading to more active star formation.  
In the face-on case, the SFR 
increases near the collisions, 
but it decreases again between the collisions down to 
the level of the isolated one. 
The SFR in the face-on disk becomes lower 
than that in the isolated disk after the sixth collision, 
when the face-on disk loses most of the cold gas 
that could later participate in star formation.

Because our simulations have not taken a cluster model 
and the associated effects into account,  
our results have limitations to be directly compared with 
the corresponding observational results 
that bear all of the complex and combined effects 
over the course of the galaxy life.  
Nevertheless, our work clearly demonstrates 
that galaxy{\textendash}galaxy hydrodynamic interactions 
can be a major mechanism 
for changing the properties of cluster LTGs, 
such as cold gas content and star formation activities. 
While very fast motion in general of the individual cluster members 
weakens the effects of the tidal interactions between them, 
it strengthens the effects of the hydrodynamic interactions. 
Besides, the frequent encounters between members  
can make the hydrodynamic interactions more important.

As mentioned in the introduction, 
some cluster LTGs showing imprints of strong  
galaxy{\textendash}galaxy hydrodynamic interactions 
have also been observed.  
Some of the cluster LTGs presented in 
\citeauthor{Ebeling+2014}$\,$(\citeyear{Ebeling+2014}; see their Figure~2)  
might have been influenced by the hydrodynamic interactions 
with neighboring galaxies 
as well as the cluster, 
because the gas tails of the LTGs 
extend to rather random directions uncorrelated with 
the deduced projected velocity vectors of the galaxies. 
In addition, NGC 4438, a highly disturbed LTG in the Virgo cluster, 
would be a good candidate showing the observational signatures 
of strong hydrodynamic interactions 
with the neighboring M86, a bright elliptical galaxy having a hot gas halo.   
The X-ray-emitting gas plume 
detected between NGC~4438 and M86 
and the spatially coinciding filaments of H$\alpha$ emission 
support the idea (\citealp{Ehlert+2013}). 
Besides, \citet{Vollmer2009} showed, 
for some Virgo cluster LTGs, that 
the linear orbital segments derived from the dynamical models 
assuming a smooth, static, and spherical ICM, 
together with the ICM density distribution 
derived from X-ray observations,  
give estimates of the ram pressure that are 
about a factor of 2 
higher than those derived from the dynamical simulations for 
NGC~4501, NGC~4330, and NGC~4569.   
\citet{Vollmer2009} also showed that 
compared to these galaxies, 
the two LTGs near M86, NGC~4388 and NGC~4438,  
require a still 2 times higher peak ram pressure than
expected from a smooth and static ICM, assuming an even 
higher stripping efficiency and/or ICM density. 
We also argue that, by taking the effects of the hydrodynamic interactions 
with neighboring galaxies into account,  
the discrepancy could be solved.

We end this work by emphasizing  
the importance of galaxy{\textendash}galaxy hydrodynamic 
interactions in order to better understand 
the distinctive properties of the galaxies in the cluster environment, 
such as the morphology-radius or morphology-density relation.  
From a numerical aspect, 
more high-quality hydrodynamic simulations 
resolving both galactic and intracluster gas 
in comparable resolutions would be more effective to unveil 
the evolution of the galaxies in a cluster.  
Simulations with well-constrained ICs 
considering the cluster 
and all of the members orbiting in it together 
would be most desirable.


\begin{acknowledgments}
We thank the anonymous referee very much for providing valuable 
comments that improved this paper. 
We also appreciate Joshua E. Barnes making the ZENO code available
and Volker Springel for providing us with GADGET-3. 
We thank the Korea Institute for Advanced Study for providing computing
resources (KIAS Center for Advanced Computation Linux
Cluster System) for this study. 
J-SH is grateful to Sungwook E. Hong for discussion 
using the Horizon Run~4 simulation data 
and to Curtis Struck, Juhan Kim, 
Maurice H.~P.~M. van Putten, 
B. L'Huillier, O.~N. Snaith, 
and S.-h. Nam  
for helpful comments.   
J-SH is supported by 
the Basic Science Research Program through 
the National Research Foundation of Korea (NRF), funded by 
the Ministry of Education (grant No. 2015R1D1A1A01059148). 
\end{acknowledgments}


\begin{center}
\MakeUppercase{appendix a\\}
\MakeUppercase{Evolution of the models in isolation}
\end{center}

\begin{figure*} [!hbt]
\centering
\includegraphics[width=16.5cm]{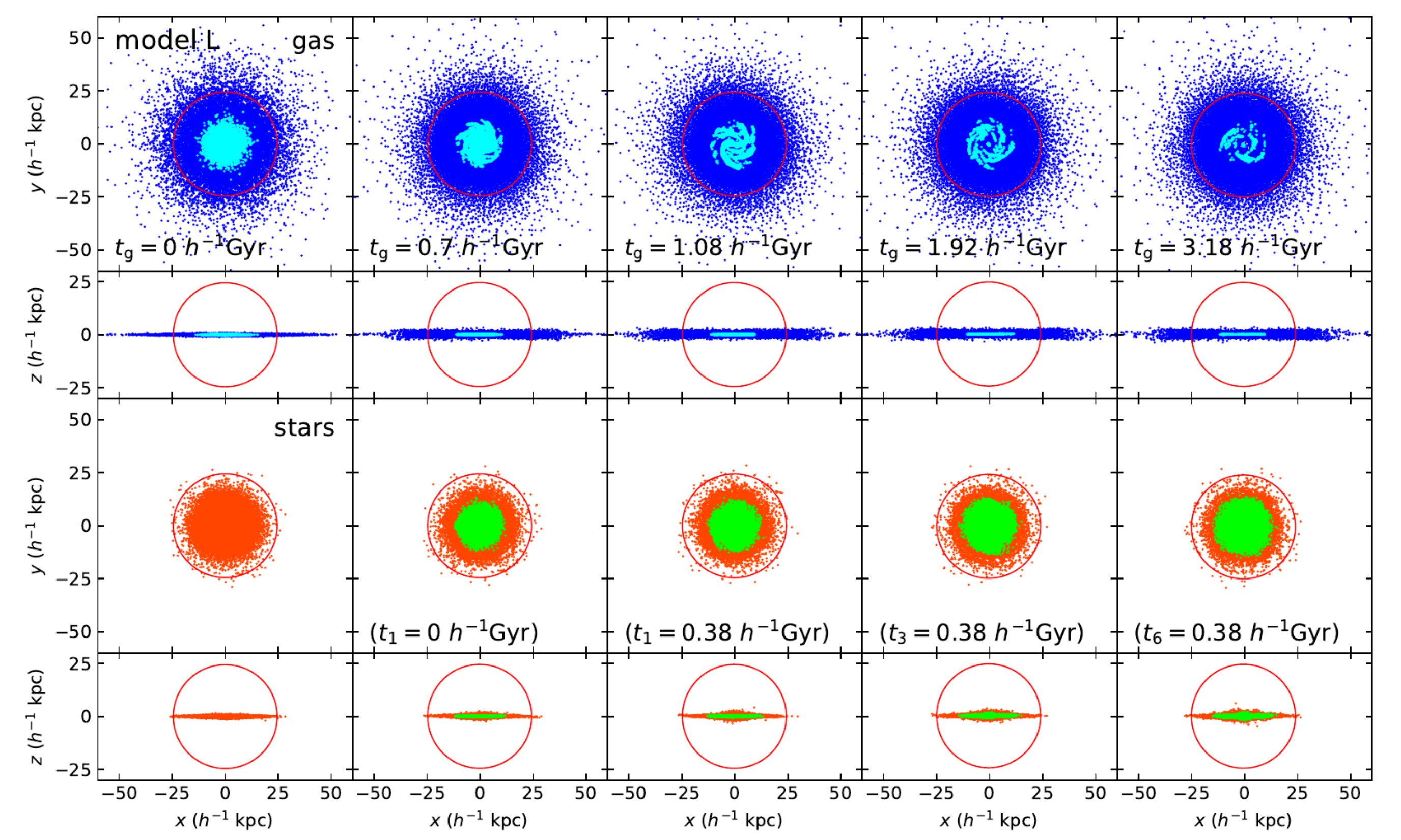} 
\caption{Five snapshots of the distribution of the disk particles 
of the isolated model~L 
at $t_{\rm g}$ = 0, 0.7, 1.08, 1.92, and 3.18~$h^{-1}$~Gyr. 
The gas and stellar particles 
are displayed with different colors in the same way as in Figure~5: 
cold non-star-forming gas in blue, star-forming gas in cyan, 
old disk stars in orange, and stars newly formed out of the gas in green. 
The large circle shown in each panel is the cross section of the sphere 
as in Figure~5, which  
encloses 90~\% of the total disk gas in mass at $t_{\rm g}$ = 0.7~$h^{-1}$~Gyr. 
The times of the second through fifth snapshots correspond to 
$t_1$ = 0, $t_1$ = 0.38, $t_3$ = 0.38, and $t_6$ = 0.38~$h^{-1}$~Gyr, 
respectively (written in parentheses in the third row; refer to Table~2).  
}
\end{figure*}

We run each of our models in isolation over 5~$h^{-1}$~Gyr 
and check the stability. 
The total energy and angular momentum of the models  
are well conserved, 
particularly with almost no change since the first few hundred Myr 
through the end of the runs.  
In Figure~8, we show the evolution of the star and gas disks of 
the LTG model~L, where more interesting phenomena occur on the disks, 
in contrast to the ETG model~EH.  
The star-forming gas (cyan) subsequently 
turns into stars (green), and  
the spiral arms and the bar develop on the disks. 
To minimize any initial fluctuations in our models, 
both models~L and EH at the time of the second snapshot 
are included in the ICs
of our encounter simulations.


\begin{center}
\MakeUppercase{appendix b\\}
\MakeUppercase{Deprojection method}
\end{center}

We adopt the geometrical deprojection algorithm 
of \citet{McLaughlin1999} to obtain an approximate 3D distribution 
of the 209 galaxies in the 
galaxy catalog of the Coma cluster (H. S. Hwang et al. 2018, in preparation). 
The geometrical technique makes only one assumption of 
circular symmetry, 
in contrast to an Abel integral, 
which requires the appropriate fitting function 
for the density of the cluster. 
We calculate the average volume densities ($n_{\rm{cl}}$) for 
a number of concentric spherical shells, 
which are divided along the 3D radii of the cluster as follows.

First, we count the number of member galaxies in every  
interval, $dR$, of the 2D projected clustercentric radius $R$.  
As illustrated in Figure~11 in \citet{McLaughlin1999}, 
let the projected radius at each boundary of the cylindrical bins 
be labeled as $R_0$, $R_1$, $R_2$, ..., $R_m$ from the center 
to the outer edge. 
(The outermost $R$ in the figure is labeled $R_2$, 
which corresponds to $R_m$ in our description.)  
We also divide the cluster into spherical shells 
at every $dr$ 
of the 3D deprojected clustercentric radius $r$ 
and label them $r_0$, $r_1$, $r_2$, ..., $r_m$.     
Choosing the same value for both $dR$ and $dr$ makes 
$r_0 = R_0$, $r_1 = R_1$, and so on.

For the outermost cylindrical bin $R_{m-1} \leq R \leq R_m$, 
because all of the galaxies observed within the cylinder   
should be located in the outermost spherical shell $r_{m-1} \leq r \leq r_m$, 
the deprojected volume density $n_{\rm cl}$ at the outermost shell 
can easily be calculated as 
the number count at the cylinder $R_{m-1} \leq R \leq R_m$  
divided by the volume of the shell $r_{m-1} \leq r \leq r_m$, 
which is intersected by the cylinder $R_{m-1} \leq R \leq R_m$ 
(i.e., the hatched regions 
between $R_{1}$ and $R_2$ in the figure).
Moving inward to the second outermost cylinder $R_{m-2} \leq R \leq R_{m-1}$, 
the number count at the cylinder 
includes the contributions from both the outermost spherical shell 
and the second outermost shell. 
Given the volume density at the outermost shell, 
the volume density at the second outermost shell can be obtained 
by solving the generalized equation~A2 of \citet{McLaughlin1999}. 
The volume density at all other shells, from the outside in, 
can also be calculated by using the same equation.

We try several different values for $dR$, 
ranging from 0.1 to 0.5~$h^{-1}$~Mpc.  
Some of them result in negative values of $n_{\rm cl}$ or very small values  
for certain spherical shells due to 
the limited number of sample galaxies. 
Choosing $dR$ = 0.3~$h^{-1}$~Mpc, 
we get all positive and reasonably smooth values for $n_{\rm cl}$.  
The obtained average number densities 
at the innermost shell through the outermost one are 
37, 29, 31, 20, 41, 32, and 19, respectively.


\begin{center}
\MakeUppercase{appendix c\\}
\MakeUppercase{Resolution test}
\end{center}

\begin{figure} [!hbt]
\centering
\includegraphics[width=7.5cm]{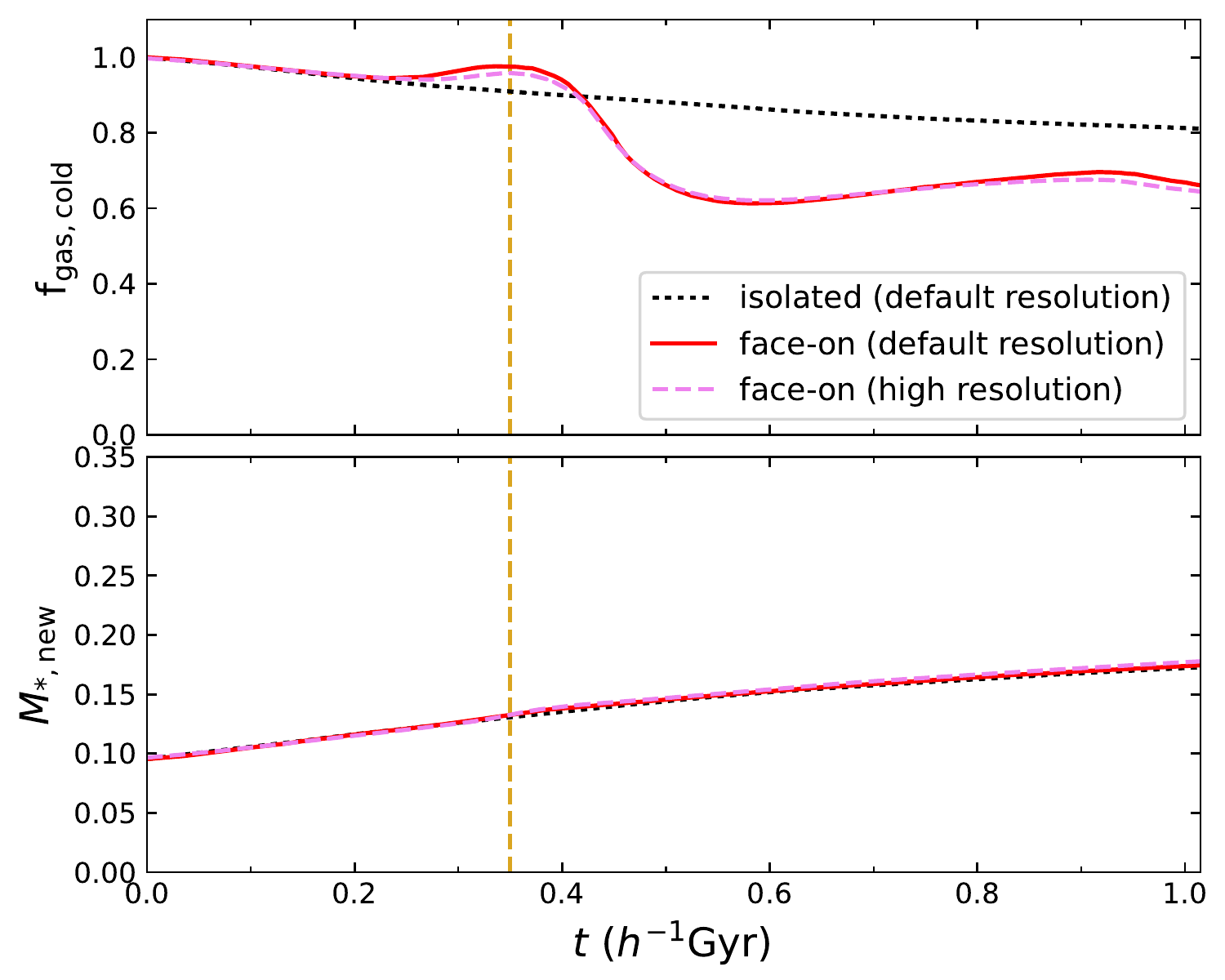} 
\caption{
Time evolution of the cold gas fraction (top panel) 
and the total mass of the newly formed stars (bottom panel) 
on the disk of the LTG model from the resolution test runs. 
The disk quantities (as in Figure~7 (c) and (h))  
are measured within the spherical volume, which had initially enclosed 
90\% of the disk gas (as drawn in Figure~5). 
The stellar mass is in units of $10^{10}\,h^{-1}\,{M_{\odot}}$.  
The values obtained from the default and high-resolution encounter runs 
are displayed with solid and dashed lines, respectively.  
Those obtained from the isolated case with the default resolution  
are shown with dotted lines for comparison.    
The vertical dashed line indicates the closest approach time 
between the LTG and ETG models, $t$ = 0.35~$h^{-1}$~Gyr. 
}
\end{figure}

In order to check whether our simulation results  
are robust to changes in resolution, 
we build our LTG and ETG models with higher resolution as well, 
using four times the number of particles for all components 
as those listed in Table~1 (``default" resolution) and  
keeping all other parameters fixed. 

As the resolution test, 
we run an LTG-ETG encounter simulation twice, 
first with the default resolution models 
and then with the high-resolution models. 
The initial configuration of the LTG and ETG models is as follows.   
Model~L is initially placed 
at ($x_{0}$, $y_{0}$, $z_{0}$) = ($-$545, 38, 0)~$h^{-1}$~kpc 
and flies face-on with a velocity 
of ($v_{x0}$, $v_{y0}$, $v_{z0}$) = (1500, 0, 0)~km~s$^{-1}$; 
model~EH is initially positioned at the origin with zero velocity. 
After 0.35~$h^{-1}$~Gyr (since the start of each run), 
models~L and EH encounter 
most closely at a distance of 35~$h^{-1}$~kpc.

In Figure~9, we show the time evolution of 
the cold gas fraction (top panel) and 
the total mass of the newly formed stars (bottom panel) 
from the default and high-resolution runs.  
We find that the values match each other well overall, and  
the main results of our simulations 
remain consistent in the different resolution runs.


\begin{center}
\MakeUppercase{appendix d\\}
\MakeUppercase{Gas content of the LTG in different situations}
\end{center}

We perform three runs, ``D", ``Cv", and ``Cd", 
to compare the evolution of the gas content of an LTG  
encountering an ETG   
at the closest approach distance of 
35~$h^{-1}$~kpc.  
(i) Run~D is the default resolution run that is described in Appendix~C. 
(ii) Run~Cv is the same as run~D, except with the initial velocity 
of model~L. 
It is $v_{x0}$ = 2500~km~s$^{-1}$ in run~Cv, 
which is about 1.7 times faster than that in run~D. 
(iii) In run~Cd, the LTG model~L collides with the different 
ETG model~``EH$^\prime$". 
The initial positions and velocities 
of both models are identical to those of run~D. 
Model~EH$^\prime$ is designed to have a gas halo 
twice as massive as that of model~EH. 
The total masses of the gas and the DM halos of  model~EH$^\prime$ 
are $M_{\rm hg} = 3.36 \times 10^{10}$ and 
$M_{\rm hd} = 164.64 \times 10^{10}\,h^{-1}{M_{\odot}}$, respectively.   
All other model parameters are the same as those of model~EH (Table~1).  
The radial density profiles of the gas halos of models~EH and EH$^\prime$ 
are presented in Figure~10.

The ram-pressure force exerting on the gas disk of the LTG, 
which flies face-on through the hot halo of the ETG,  
is proportional to the mass density of the gas halo 
and the square of the relative velocity of the LTG 
with respect to the ETG. 
Thus, the strength of the ram pressure on the gas disk   
is expected to be strongest in run~Cv, second-strongest in run~Cd, 
and weakest in run~D. 
Figure~11 presents the evolution of the cold gas fraction of model~L 
from the three comparison runs. 
The cold gas (cold non-star-forming gas + star-forming gas) fraction 
of model~L becomes the lowest in run~Cv after the collision 
because of the strongest ram pressure. 
After the cold gas fraction reaches the minimum, it rises  
as some of the cold gas that is not completely stripped off the disk 
falls back onto the disk. 
This trend appears more strongly in runs~Cd and D than in run~Cv.   
The ram pressure on the disk also leads 
the rise of the star-forming gas fraction near the collision, 
compared with that of the isolated case.

\begin{figure} [!hbt]
\centering
\includegraphics[width=6.0cm]{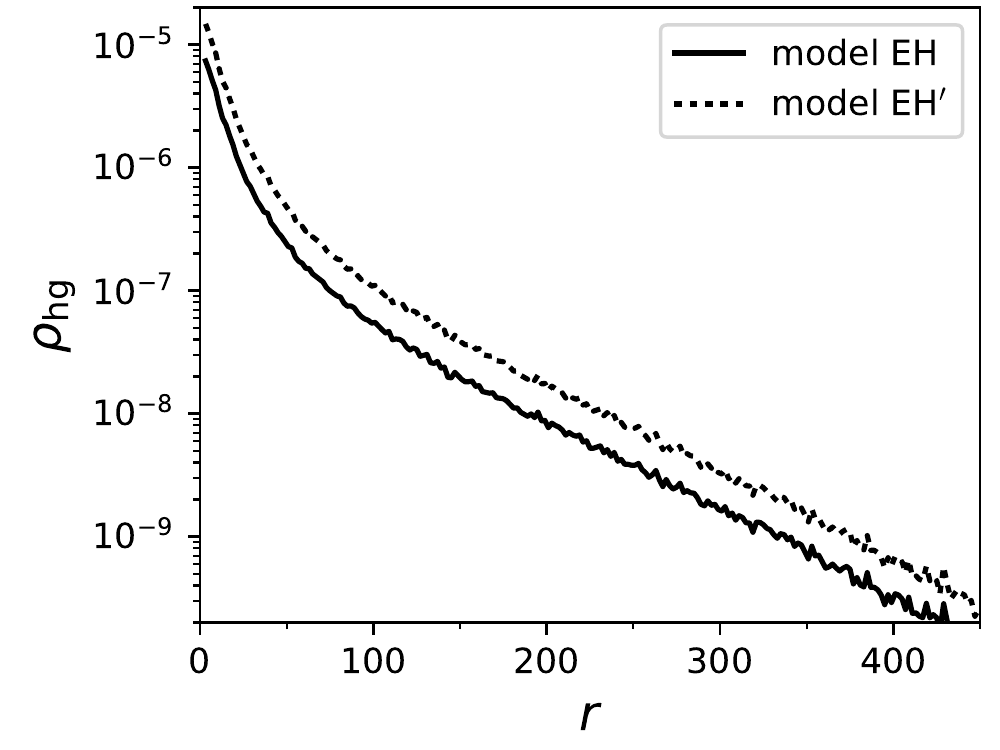} 
\caption{
Spherically averaged density profile of 
the gas halo of model~EH$^{\prime}$ (dotted line) 
in comparison with that of model~EH (solid line).  
The gas density $\rho_{\rm{hg}}$ is in units of 
$10^{10}\,h^{2}{M_{\odot}}\,{\rm{kpc}}^{-3}$, 
and the spherical radius $r$ is in $h^{-1}$~kpc. 
}
\end{figure}

\begin{figure} [!hbt]
\centering
\includegraphics[width=7.5cm]{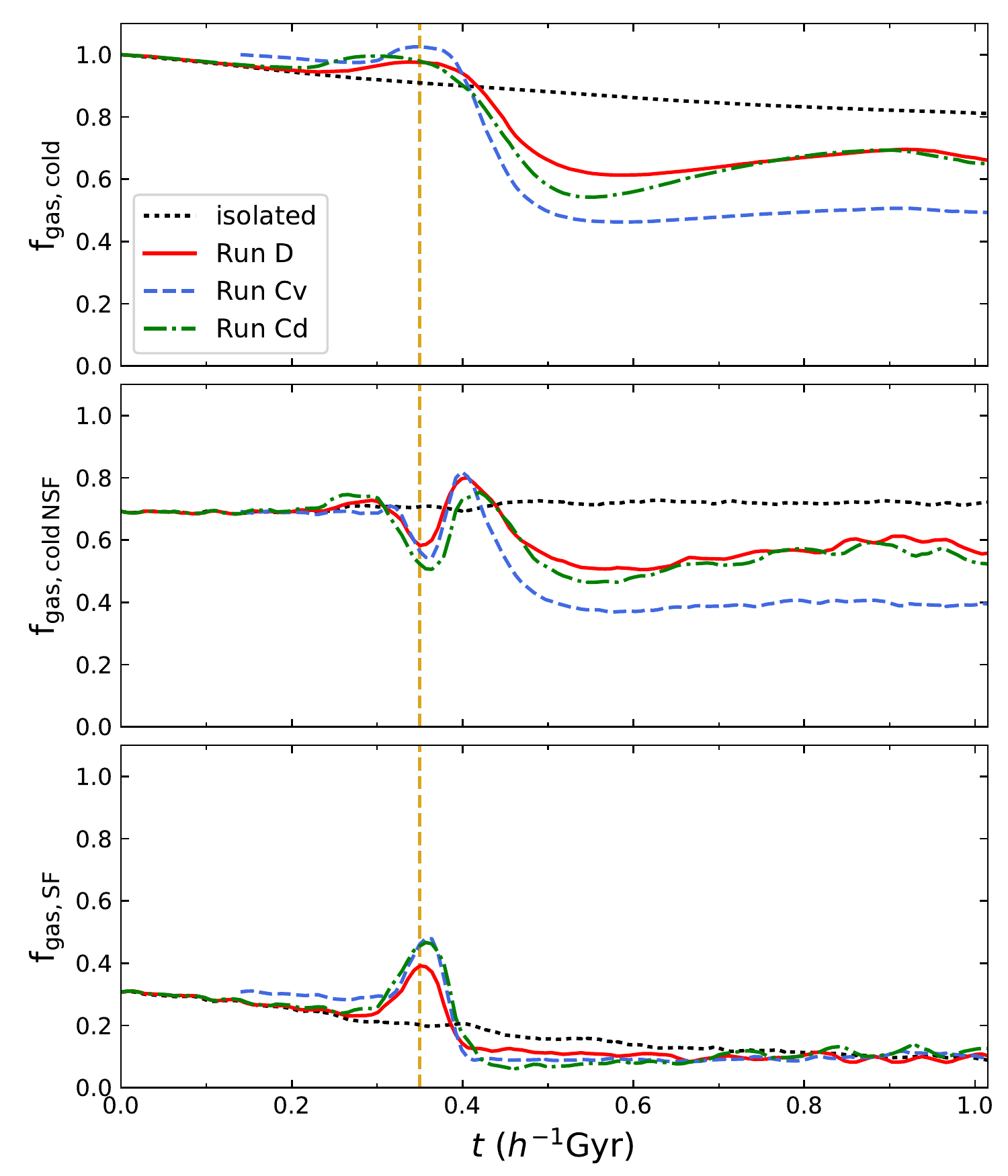} 
\caption{
Time evolutions of the cold gas  
(cold non-star-forming gas + star-forming gas) fraction (top panel), 
cold non-star-forming gas fraction (middle panel), 
and star-forming gas fraction (bottom panel)  
from the three comparison runs~D, Cv, and Cd 
(solid, dashed, and dot-dashed curves, respectively). 
The dashed curves for run~Cv are shifted along the $x$-axis 
toward the right by 
1.4~$h^{-1}$~Gyr so that the closest approach time between 
the LTG and the ETG models in run~Cv matches with those in runs~D and Cd, 
which is 0.35~$h^{-1}$~Gyr (vertical dashed line).  
}
\end{figure}


\end{document}